# A Novel Conditional Adversarial Diffusion Model for High-Fidelity OCT-to-OCT Image Translation and Precise Anti-VEGF Therapy Response Prediction in Diabetic Macular Edema

Reza Rasti, *Member, IEEE*, Armin Biglari, Sobhan Mohammadi Hafshejani, Alireza Dehghani, Michael J. Allingham, and Sina Farsiu, *Fellow, IEEE*

***Abstract*— Diabetic macular edema (DME) is a leading cause of severe vision loss among working-age adults. Although intravitreal anti-vascular endothelial growth factor (anti-VEGF) injections currently serve as the first-line therapy for DME, a notable subset of patients fails to respond despite the substantial cost and clinical burden, suggesting they may benefit more from alternative interventions. Automated, noninvasive pre-treatment response assessment could therefore spare non-responders from the current trial-and-error treatment paradigm. In this study, we propose a novel conditional adversarial diffusion framework that synthesizes post-treatment optical coherence tomography (OCT) images from baseline scans, enabling the derivation of each patient's initial response to anti-VEGF therapy. The model was developed with subject-level 5-fold cross-validation on a US cohort (n = 127) and externally validated on an independent Iranian cohort (n = 44). Benchmarked against fourteen baseline and state-of-the-art translation models, the proposed model achieved the best treatment-response prediction performance among the methods evaluated for that task on both cohorts, with balanced accuracies of 85.7% and 93.0% on the US cohort and 85.2% and 95.3% on the external Iranian cohort at the 10% and 20% central subfield thickness reduction thresholds, respectively. Ultimately, this framework offers a decision-support tool to help ophthalmologists anticipate initial anti-VEGF outcomes in DME prior to treatment initiation, toward personalized therapy selection. The source code will be made publicly available upon acceptance.**

## I. Introduction

DIABETIC macular edema (DME), caused by alterations in retinal microvasculature and other retinal cell types, is a major cause of severe vision loss in individuals with diabetes [1]. Intravitreal injection of anti-vascular endothelial growth factor (anti-VEGF) agents are considered the first line of therapy for DME [2]. However, between 31% and 65% of patients treated with anti-VEGF therapy exhibit an incomplete clinical response, defined as persistent vision loss or retinal swelling, and would benefit from alternative therapies [3]. Additionally, the frequent anti-VEGF injections often required for effective treatment of DME are costly and carry the risk of potentially severe side effects [3], [4]. Thus, there is significant interest in predicting the efficacy of anti-VEGF therapy prior to injection to determine the most favorable treatment regimen for each individual.

Optical coherence tomography (OCT) is the most widely utilized imaging modality in ophthalmology, playing a central role in both the initial diagnosis of retinal disorders and the ongoing assessment of therapeutic outcomes. In the context of anti-VEGF therapy, where treatment response in conditions like DME is inherently multifactorial, OCT has become the clinical benchmark for monitoring efficacy. Its non-invasive capacity to capture quantitative anatomical indicators, particularly central subfield thickness (CST) and intra-/subretinal fluid volumes, renders it indispensable for informing retreatment strategies and individualizing patient care [5].

Many machine learning (ML) and deep learning (DL) algorithms have been successfully applied for a broad range of OCT image analysis tasks [4], [6], [7]. Artificial intelligence (AI) can employ various strategies to evaluate large, complicated datasets to uncover new information that cannot be obtained through expert manual analysis [8], shaping a new landscape for ocular disease diagnosis and prognosis. An exciting application of AI-based OCT image analysis is predicting the efficacy of anti-VEGF therapy in managing retinal diseases, including DME. It has been hypothesized that the OCT volumes contain subtle, complex patterns and textural

*This work was supported by the Department of Biomedical Engineering, University of Isfahan. (Corresponding author: *Reza Rasti*)

R. Rasti and A. Biglari contributed equally to this work and are co-first authors.

R. Rasti, A. Biglari, and S. M. Hafshejani are with the Department of Biomedical Engineering, Faculty of Engineering, University of Isfahan, Isfahan, Iran (e-mails: r.rasti@eng.ui.ac.ir; armin.biglari@mehr.ui.ac.ir; s.mohammadi@mehr.ui.ac.ir). R. Rasti is also a research collaborator with the VIP Lab, Department of Biomedical Engineering, Duke University, Durham, NC 27708, USA (e-mail: reza.rasti@duke.edu).

A. Dehghani is with the Isfahan Eye Research Center, Department of Ophthalmology, Isfahan University of Medical Sciences, Isfahan, Iran (e-mail: dehghani@med.mui.ac.ir).

M. J. Allingham is with the Department of Ophthalmology, Duke University School of Medicine, Durham, NC 27708, USA (e-mail: mike.allingham@duke.edu).

S. Farsiu is with the Department of Biomedical Engineering, Duke University, Durham, NC 27708, USA. S. Farsiu is also with the Department of Ophthalmology, Duke University Medical Center, Durham, NC 27708, USA (e-mail: sina.farsiu@duke.edu).

signatures that are uniquely suited for AI models to learn latent computational biomarkers, thereby capturing an integrated signature of the disease state that can predict future treatment response [4]. Section II reviews relevant ML and novel AI technologies that have been proposed in recent years. Despite promising results, the accuracy of these techniques is far from perfect.

This study proposes the source-conditional OCT diffusion generative adversarial network (SCOD-GAN), a novel conditional diffusion GAN model developed to synthesize post-treatment OCT scans of anti-VEGF-treated DME patients from baseline OCT characteristics, from which patient responses to therapy can be derived as either responsive or non-responsive. This method is a critical step in a continuum of research aimed at developing an automated personalized medicine tool that assists ophthalmologists in predicting the responses of DME patients to available treatments and subsequently selecting the most effective line of therapy for each patient. The contributions of the current study are as follows:

(1) A novel conditional diffusion-GAN method is introduced for high-fidelity OCT-to-OCT image translation. The method employs a cascaded denoising diffusion-GAN architecture in which conditional GANs parametrize the reverse process of the diffusion model. Through this synergistic integration, stable convergence and high-quality synthesis were achieved within only 4 sampling steps.

(2) Based solely on pre-treatment OCT scans, the proposed thickness-aware SCOD-GAN framework generates anatomically consistent post-treatment OCT images from which anti-VEGF treatment-response predictions can be derived for DME patients. The pipeline follows a two-stage cascade: pre-treatment OCTs are first mapped into anatomically informed retinal thickness masks, which subsequently guide post-treatment OCT synthesis. Rather than relying on a dedicated neural classification head, treatment-response status is derived in a post-generation step by measuring the CST change between the pre-treatment and synthesized post-treatment OCTs. Joint optimization across the cascaded image-generation objectives allows the diffusion and GAN modules to co-adapt during training, capturing therapy-induced anatomical changes while retaining patient-specific retinal features. Ultimately, this explicit structural conditioning provides an interpretable representation of the modeled anatomical changes, facilitating downstream OCT-derived response assessment in personalized care.

(3) Advanced training strategies are incorporated into SCOD-GAN. A compound loss formulation with task-specific weighting is employed to balance anatomical fidelity and image realism, while hybrid supervision is introduced in the first-stage generator to enforce structural correctness. This anatomical guidance is then transferred to the second stage, ensuring clinically meaningful synthesis of post-treatment OCTs.

(4) The efficiency of the proposed model was evaluated through extensive experiments using two distinct spectral domain OCT datasets: the Duke-DME dataset (introduced in [4]) and a new Isfahan-DME dataset. Comprehensive ablation and sensitivity analyses confirmed the indispensability of the proposed cascaded design, compound loss, and hybrid supervision. Superior performance over state-of-the-art methods was achieved across multiple metrics, while maintaining computationally efficient inference.

(5) The newly curated Isfahan-DME retinal OCT dataset, specifically designed for predicting anti-VEGF treatment outcomes in DME, is made available upon academic request to facilitate future predictive modeling research.

The remainder of this paper is as follows: Section II reviews studies on predicting the anti-VEGF therapeutic response in patients with retinal disorders. Section III introduces our proposed methods, including the SCOD-GAN model architecture. Section IV details the study's data, evaluation metrics, implementation process, and experimental results. Sections V and VI provide a discussion and conclusion of the study, along with suggestions for future research directions.

## II. Related Work

### A. Non-Generative ML and DL Methods

Early studies using invasive imaging systems such as fluorescein angiography combined with semi-automatic segmentation showed the possibility of predicting anti-VEGF treatment response in DME subjects [9]. The predictive value of OCT image analysis using ML and DL to assess the efficacy of anti-VEGF agents has been exploited for DME (Table I), age-related macular degeneration (AMD), and retinal vein occlusion (RVO) [10], [11]. In 2018, Chen *et al.* used artificial neural networks (ANNs) for a regression problem to predict the final visual acuity (VA) of DME patients with frequent injections of Ranibizumab [12]. Complementary studies were later conducted by Rasti *et al.* and Roberts *et al.,* utilizing their convolutional neural networks (CNNs)-based frameworks to analyze structural changes in the retina after the treatment process [4], [13]. Cao *et al.* categorized the response of patients with DME to anti-VEGF treatment using random forest (RF) and support vector machines (SVMs). This was done by analyzing features derived from the segmentation of retinal layers and disruption ratio, segmentation of intraretinal and subretinal fluid, and the count of hyper-reflective foci [14]. Alryalat *et al.* used a modified UNet model to first segment retinal fluids, hyper-reflective foci, and disruption of the ellipsoid zone before classifying patients into responsive and non-responsive classes using a DL-based model [2]. Liu *et al.* employed an ensemble of DL-based models to predict both central foveal thickness (CFT) and best-corrected visual acuity (BCVA) in patients receiving anti-VEGF therapy. This approach integrated OCT image features with relevant clinical parameters [15]. Patel *et al.* evaluated four benchmark ML regression models in predicting post-treatment CST and distant vision BCVA utilizing a set of baseline clinical parameters [16]. Relevant studies by Zhang *et al.* and Ye *et al.* assessed the therapeutic response of DME patients to the treatment [3], [17]. Gallardo *et al.* further employed two RF models to forecast long-term treatment requirements in patients diagnosed with neovascular AMD (nAMD), RVO, and DME [18].

TABLE I
COMPREHENSIVE OVERVIEW OF RESEARCH STUDIES ON PREDICTIVE MODELS FOR ANTI-VEGF TREATMENT EFFICACY IN DME PATIENTS

| Author | Target | Data | Task | Input(s)* | Method | Evaluation Metrics* |
|---|---|---|---|---|---|---|
| Chen *et al.*, 2018 [12] | Final visual acuity of DME patients treated with ranibizumab at 52, 78, or 104 weeks | 512, 483, and 464 eyes that received treatment at 53, 78, and 104 weeks, respectively | Regression | Age, Sex, Type of diabetes, use of Insulin, Hypertension and Hypercholesterolemia under treatment, HbA1c level, Lens status, Degree of DR severity, Baseline macular OCT value (including central point, central, inner and outer superior/ nasal/ inferior/ temporal part), Baseline VA, and Timetable of ranibizumab treatment | ANN | The respective correlation coefficient for training, testing, and validation groups: 0.75, 0.77, and 0.70 (52 weeks); 0.79, 0.80, and 0.55 (78 weeks); 0.83, 0.47, and 0.81 (104 weeks) |
| Rasti *et al.*, 2020 [4] | Responsive or non-responsive DME patients treated with Ranibizumab, Aflibercept, and Bevacizumab | 127 patients who received treatment in 3 consecutive injections | Prediction | First-visit SD-OCT ROI images of DME patients | CNN-based CADNet | AUC of 0.866, average precision, sensitivity, and specificity of 85.5%, 80.1%, and 85.0% |
| Cao *et al.*, 2020 [14] | The anti-VEGF therapeutic response of DME patients | 712 SD-OCT B-scans of patients before and after one month of treatment | Classification | Image features, including retinal layer segmentation and disruption ratio, Intraretinal and subretinal fluid segmentation and area quantization, the count of HRF dots, and Optical density ratio of IRF and SRF obtained by DL methods | RF, SVM | Best performance: 90.7% specificity, 87.7% sensitivity, and 95.1% AUC from the RF classifier |
| Roberts *et al.*, 2020 [13] | The volumetric change of IRF and SRF of DME patients during anti-VEGF treatment | SD-OCT images and clinical information that were assessed every 4 weeks in 570 patients with a 1-year follow-up | Regression, Statistical Analysis | Calculated the volumes of the IRF and the SRF for both the central fovea and the parafovea. Prior classification of voxels as background, IRF, or SRF in SD-OCT scans was performed after a preprocessing step consisting of automatic alignment and registration using a CNN approach. | Linear Mixed Effects Model | Baseline SRF presence correlated with poorer BCVA scores (63.2 vs 66.9 ETDRS scores), but also with a greater improvement in BCVA during follow-up (4 scores per 4 weeks vs 0.4 scores). |
| Liu *et al.*, 2021 [15] | The post-treatment central fovea thickness and best corrected visual acuity values | A total of 363 OCT images and 7,587 clinical data records from 363 eyes receiving 3 monthly injections | Regression | Deep fusion features extracted from the OCT images using the ensembled DL models and combined with clinical parameters | Classical ML | Ensembled system's performance in external validation: MAE, RMSE, and $R^2$ of 68.08, 97.63, and 0.74, respectively, for CFT prediction and 0.13, 0.2, and 0.68, respectively, for BCVA prediction. |
| Alryalat *et al.*, 2022 [2] | Good or poor response to intravitreal anti-VEGF injections among DME patients. | Single B-scans centered at the fovea from both the pre-treatment and post-treatment OCT images of 101 patients (60 good and 41 poor responders) | Classification | Segmented images by a novel UNet model using Inner/Outer IRF, SRF, HRF, and Disruption of the ellipsoid zone for annotation | EfficientNet-B3 | Segmentation accuracy of 95.9%, with a specificity of 98.9%, and a sensitivity of 87.9%. The classification accuracy reached 75%, comparable to that of ophthalmologists. |
| Zhang *et al.*, 2022 [3] | VA at 1 month after anti-VEGF injection in DME patients with 1 + PRN treatment regimens | VA and OCT of pre- and postoperative (1 month) data from a total of 281 eyes | Regression | Medical records such as age, gender, VA baseline, and VA 1-month Clinical features: DRT, CME, SRD, PHT, TRD, DRIL, HRF, etc. | LR, SVM, KNR, RF, Ridge Regressor, LR + RF | MAEs of VA predictions=0.137-0.153 logMAR (within 7-8 letters), and the MSEs=0.033-0.045 logMAR (within 2-3 letters) for the one-month VA predictions, respectively. |
| Xu *et al.*, 2022 [19] | Post-treatment OCT to predict the short-term response to anti-VEGF therapy in DME patients | A total of 561 pre- and post-treatment + 71 post-treatment OCT images | Image-to-Image Translation | Pairs of pre- and post-treatment OCT images as GAN models' input | Pix2pix | MAE of the CST between the synthetic images and the actual OCT images=24.51±18.56 $\mu m$ |
| Patel *et al.*, 2022 [16] | Possible post-treatment CST value and DV BCVA in eyes with DME | A total of 1874 samples (OCT images + clinical parameters) from 906 patients | Regression | Clinical and structural parameters of near BCVA value, DV BCVA value, number of previous injections, number of days after the injection, ETDRS grid subfields, pre-CST, and pre-subfield average volume | SVR, RF, RBF | Linear SVR outperformed other models for predicting both CST and DV BCVA. Correlation coefficients for CST ranged from 0.65 to 0.85, while DV BCVA achieved even stronger correlations (0.83 to 0.92). |
| Ye *et al.*, 2023 [17] | Fundus structural and functional parameters and treatment plan in 6 months after different anti-VEGF treatments | 254 DME patients and 2,763 OCT images from three countries | Classification, Regression | Eight parameters, including age, gender, pre-VA, pre-CRT, pre-RIF, pre-SRF, pre-PED, and pre-HRF, at baseline. | RF, GRNN, PNN, ELM, SVM, OLS, BPN, RBF, SVR | The model achieved near-perfect accuracy in predicting RPE detachment (AUC close to 1.000). Visual acuity forecasts had an average error of 0.3 to 0.4 logMAR. The model accurately predicted the need for continuous injection in 70% of cases. |
| Liu *et al.*, 2023 [20] | Post-treatment OCT to predict the response of DME patients to anti-VEGF therapy | 715 and 103 pairs of pre-and post-treatment OCT images of patients | Image-to-Image Translation | Pairs of pre- and post-treatment OCT images as GAN models' input | CycleGAN, PairGAN, Pix2pixHD, RegGAN, SPADE, UNIT | The best performance: MAE of the CST between the synthetic images and the actual OCT images= 26.74±21.28 $\mu m$. |

***Abbreviations:*** **AUC**=Area under the Receiver Characteristic Operative Curve, **BPN**=Back Propagation Network, **CME**=Cystoid Macular Edema, **CRT**=Central Retinal Thickness, **DR**=Diabetic Retinopathy, **DRIL**=Disorganization of Retinal Inner Layer, **DRT**=Diffuse Retinal Thickening, **DV**=Distance Vision, **ELM**=Extreme Learning Machine, **ETDRS**=Early Treatment Diabetic Retinopathy Study, **GRNN**=Generalized Regression Neural Network, **HbA1c**=Glycated Hemoglobin, **HRF**=Hyper-reflective Foci, **IRF**=Intra-retinal Fluid, **KNR**=K Neighbors Regressor, **LR**=Linear Regression, **OLS**=Ordinary Least Squares, **PED**=Pigment Epithelial Detachment, **PHT**=Posterior Hyaloid Traction, **PNN**=Probabilistic Neural Network, **PRN**=Pro re nata, **RBF**=Radial Basis Function, **RIF**=Retinal Interlayer Fluid, **RMSE**=Root Mean Square Error, **ROI**=Region of Interest, **SD-OCT**=Spectral Domain-OCT, **SRD**=Serous Retinal Detachment, **SRF**=Sub-retinal Fluid, **SVR**=Support Vector Regression, **TRD**=Traction Retinal Detachment.

## B. Image-to-Image Translation and Generative Modeling

1) Background: Pix2Pix [21] and CycleGAN [22] models laid the groundwork for image generation tasks, with Pix2Pix using a supervised UNet architecture for precise image translation, and CycleGAN excelling in unpaired image translation through its cycle-consistency approach. Building on these foundations, UNIT [23] and MUNIT [24] introduced shared latent spaces and separate style distributions, enhancing both translation accuracy and flexibility. BicycleGAN [25] models multiple plausible outputs from paired training data, while Pix2PixHD [26] extended Pix2Pix to handle high-resolution images. In more recent developments, models like HengSRC [27] have focused on preserving fine details through patch-wise contrastive learning, which is crucial for accurate medical imaging. Özbey et al. introduced the adversarial diffusion model SynDiff [28], which couples cycle-consistent adversarial learning with accelerated large-step diffusion sampling for unpaired medical image translation. Fast-DDPM [29] addresses the need for faster diffusion models, making them more practical for real-time clinical applications. Additionally, advancements in anatomically controllable and multimodal image synthesis have been made by SGD [30] and I2I-Mamba [31] models, addressing the demand for both precision and comprehensive diagnostic imaging. Lastly, selfRDB [32] has refined the image translation process through recursive diffusion, ensuring the high-quality outputs required in medical settings. The FGDM model [33] is developed to address the challenge of structural preservation in medical image translation by leveraging frequency-domain filters, ensuring critical anatomical information is retained throughout the diffusion process. PatchGCL [34] is a model developed to enhance patch-wise contrastive learning by integrating graph neural networks (GNNs) to capture topology-aware features, improving the semantic correspondence in image translation. The StegoGAN [35] model has recently been proposed to improve the semantic consistency of translated images in non-bijective image-to-image translation, eliminating the need for additional post-processing or supervision.

*2) Generative ML Approaches for Predicting Post-Treatment OCT Images:* The non-generative works above demonstrated the potential applications of ML tools in selecting treatment strategies for retinal disease and opened the door for further research in this area. Relevant studies have also been conducted to predict the post-treatment OCT images of anti-VEGF-treated patients using generative networks. In 2020, Lee *et al.* trained a conditional GAN network on augmented paired OCT B-scans originally obtained from patients with nAMD, and also a set of baseline fluorescein angiography (FA) or indocyanine green angiography (ICGA) images to generate post-treatment scans [36]. They validated its performance in predicting the presence of distinct structural parameters, achieving accuracy, specificity, and negative predictive value ranges of 80.7-96.3%, 97.3-99.0%, and 59.0-98.3%, respectively. A limitation of this approach is the invasive nature of FA and ICGA tests, which require the injection of contrast agents [37]. Liu *et al.* proposed a Pix2pixHD method to predict the short-term response of nAMD patients by synthesizing post-treatment OCT images according to the baseline images [38]. The quality, authenticity, and predictive power of the output images were validated through three experiments conducted by retinal ophthalmologists, with 92% of the output images meeting an adequate standard for additional clinical interpretation. The accuracy in predicting the macular status as wet or dry was 85%. In 2022, Xu *et al.* also used Pix2pixHD to generate post-treatment images of DME and RVO patients in two studies [19], [39]. They quantitatively evaluated the quality of generated images compared with the real post-treatment scans with the mean absolute error (MAE) of the CST, which was 24.51±18.56 µm for the DME and 26.33±15.81 µm for the RVO studies. However, the data they used in the DME study were all from the same health center, and the training and validation data were split up based on chronological order. With similar objectives, Moon *et al.* applied an attention GAN model to generate post-treatment OCT images of nAMD patients and predict their post-treatment fluid status [40]. This study used only Korean patients.

Recently, Liu *et al.* predicted post-treatment OCT images of patients with DME by applying six different GAN models (CycleGAN, PairGAN, Pix2pixHD, RegGAN, SPADE, UNIT) [20]. The best reported MAE for CST between the synthetic and real post-treatment images was 26.74±21.28 µm. This study also used single-center data with a chronology-based train and validation set separation. To prevent GAN models from ignoring local information, Zhao *et al.* presented the BAABGAN framework [41] to synthesize OCT images after nAMD treatment. Large-scale longitudinal data were obtained to train and evaluate the model with acceptable performance. However, the study did not incorporate significance testing or cross-validation techniques, which limits the generalizability of the findings across different datasets.

In this study, a novel framework is introduced to address the limitations of previous methodologies in synthesizing high-fidelity post-treatment OCT images for anti-VEGF-treated DME patients. During training, the network utilizes paired pre- and post-treatment OCT scans alongside post-treatment retinal masks; at inference, however, it requires solely the pre-treatment OCT data. By incorporating key structural characteristics of the retina, such as total retina thickness, the model improves upon the fidelity of the synthesized images, making them more representative of actual post-treatment outcomes. The model's performance was rigorously evaluated using two distinct, geographically diverse datasets, demonstrating its robustness and generalizability.

# III. Materials and Methods

## A. Datasets

This study utilizes two distinct retinal OCT datasets of DME patients. The Duke-DME dataset [4], comprising 127 cases, includes retinal OCT volumes acquired before and after a series of three anti-VEGF injections administered at intervals of 4 to 6 weeks. Each volume contains 61 B-scans with a resolution of 768×496 pixels, accompanied by binary segmentation masks that represent the total retinal thickness, spanning from the nerve fiber layer (NFL) to the retinal pigment epithelium (RPE).

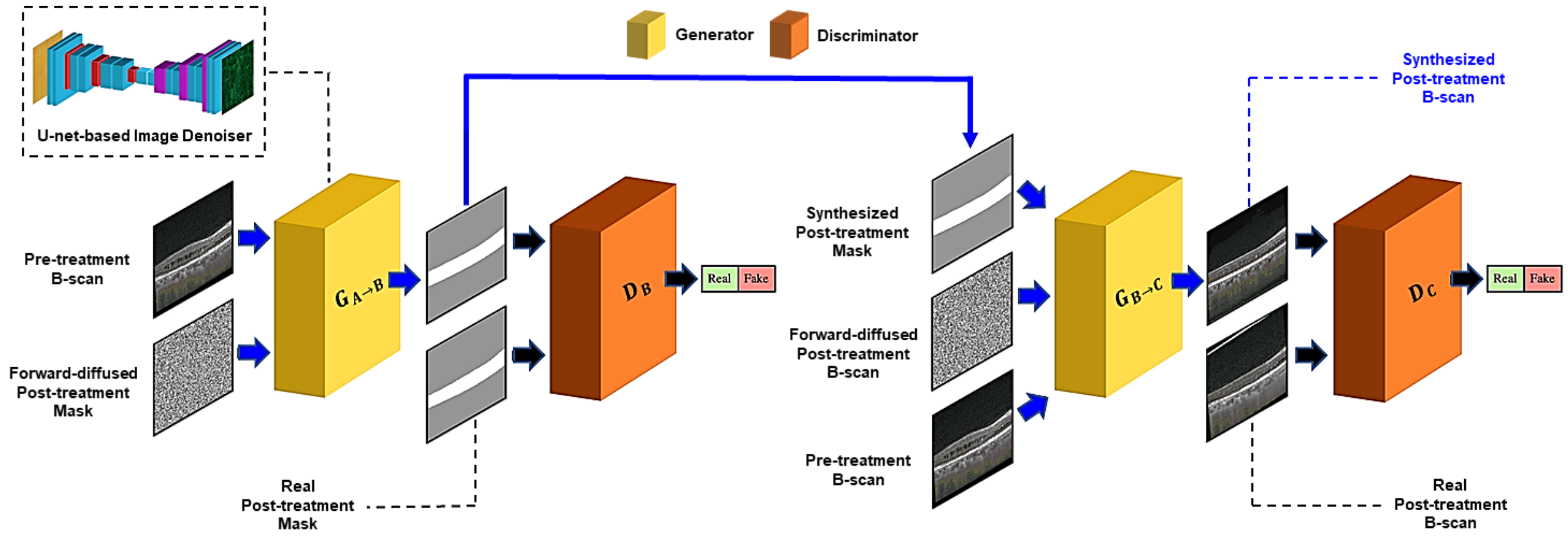

Fig. 1. Overview of the SCOD-GAN framework: The architecture consists of two main components, each with its diffusive generators and discriminators. $G_{A\to B}$ generates a post-treatment retinal OCT mask and $D_B$ evaluates its validity. $G_{B\to C}$ generates the post-treatment OCT image and $D_C$ assesses its validity. The blue paths illustrate the inference pipeline, where forward-diffused noisy images are substituted with random Gaussian noise samples of $\mathcal{N}(0, \boldsymbol{I})$. All components are jointly optimized in an end-to-end manner, enabling seamless feature sharing and coordinated learning across the stages.

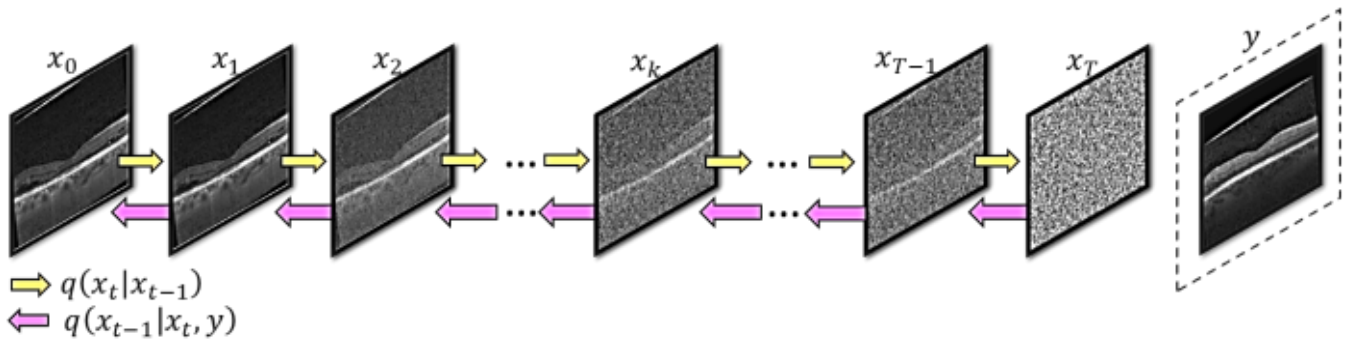

Fig. 2. Diffusion models progressively switch between samples $x_t$ via adding/suppressing noise in each forward/reverse step. Image-to-image translation involves using a source image, denoted as $y$, as a conditioning input for the reverse processes.

The second dataset, referred to as the Isfahan-DME dataset, was collected at Feiz Hospital in Isfahan and consists of 44 local DME cases treated with either bevacizumab (20 cases) or aflibercept (24 cases). Patients received 2 (n=12), 3 (n=27), or 4 (n=5) intravitreal injections spaced 4 to 6 weeks apart. For each case, OCT scans are acquired both before and after treatment, along with associated clinical data (e.g., visual acuity) and structural retinal biomarkers (e.g., central retinal thickness). The dataset includes volumes with 19 (n=9), 25 (n=23), or 31 (n=12) B-scans, each with a resolution of 768×496 pixels, except for four cases in which the B-scan size was 512×496 pixels. Similar to the first dataset, each scan includes a corresponding binary mask of the total retinal thickness. The second dataset was approved by the Institutional Review Board of Isfahan University of Medical Sciences (Protocol No. IR.ARI.MUI.REC.1402.336), in accordance with institutional ethical standards and the Declaration of Helsinki.

## B. Image Pre-processing Pipeline

For the analysis of retinal OCT data, each B-scan is geometrically normalized to ensure consistent anatomical alignment across the dataset. Each scan, along with its corresponding total retina segmentation mask, is centrally cropped and resized to 256×256 pixels to standardize the input dimensions. To account for inter-scan variability in retinal orientation, the outer boundary of the RPE is extracted. The slope between the endpoints of this boundary is used to estimate the retinal tilt angle, which is then corrected through image rotation. A subsequent vertical translation is applied to center the retinal structure within the B-scan based on the detected boundary position. This alignment procedure reduces anatomical variability and positional artifacts, thereby improving the consistency and robustness of downstream deep generative models in the retinal image analysis task.

## C. Overview of The Proposed Framework

Fig. 1 illustrates the overall architecture of the proposed SCOD-GAN OCT-to-OCT translation framework, which comprises two pairs of diffusive generators and discriminators operating in cascade. In the first stage, a UNet-based denoising generator conditioned on the real pre-treatment B-scan synthesizes the post-treatment retinal mask, while its paired discriminator learns to distinguish between ground-truth and synthetic post-treatment masks. In the second stage, the pre-treatment B-scan and the synthesized total-retinal mask are concatenated channel-wise to serve as a thickness-aware anatomical prior, guiding the second generator to synthesize the final post-treatment B-scan. Concurrently, the second discriminator differentiates real from synthesized post-treatment B-scans. In a subsequent post-generation step, each patient's therapeutic status (responsive vs. non-responsive) is derived by quantifying the relative change in CST between the pre-treatment and synthesized post-treatment OCT scans.

## D. Background: Conditional Diffusion Models

As illustrated in Fig. 2, the diffusion model transforms an image into random noise through the *forward diffusion process*. Conversely, the models employ a denoising approach to incrementally remove the introduced noise and reconstruct the image over T time steps, a procedure referred to as the *reverse denoising process*. The *forward diffusion process* involves subjecting an input image $x_0 \sim q\,(x_0)$ to Gaussian noise, thereby generating a sample $x_T$ from an isotropic Gaussian distribution for sufficiently large $T$. This process is characterized as a Markov chain, where the transition from $x_{t-1}$ to $x_t$, $t \in \{1, \dots, T\}$, follows a specified forward transition probability:

$$q(x_t|x_{t-1}) = \mathcal{N}(x_t; \sqrt{1-\beta_t}x_{t-1}, \beta_t \boldsymbol{I}), \qquad (1)$$

where $\mathcal{N}$ is the Gaussian distribution with mean $\sqrt{1-\beta_t}\, x_{t-1}$ and variance $\beta_t \boldsymbol{I}$. The dimensions of the identity matrix $\boldsymbol{I}$ match that of $x_0$, and $T$ represents the total number of diffusion steps. Within each transition of the Markov chain, the Gaussian noise is progressively incorporated, which results in a sequence of noise samples, denoted as $x_1, \dots, x_T$.
Since the samples at each step $t$ depend solely on the samples from the previous moment $t-1$, under the assumption that $t$ follows a uniform distribution:

$$q(x_t|x_0) = \mathcal{N}\left(x_t; \sqrt{\bar{\alpha}_t}x_0, (1-\bar{\alpha}_t)\boldsymbol{I}\right), \quad (2)$$

where $\alpha_t = 1-\beta_t$ and $\bar{\alpha}_t = \prod_{i=1}^{t}\alpha_i$. Parametric reorganization leads to the following process:

$$x_t = \sqrt{\bar{\alpha}_t}\, x_0 + \sqrt{1-\bar{\alpha}_t}\, \varepsilon, \quad \varepsilon \sim \mathcal{N}(0, \boldsymbol{I}), \quad (3)$$

which for sufficiently large $T$, $x_T$ will be pure Gaussian noise.

In the *reverse denoising process*, gradual denoising of samples is performed through a Markov chain from $x_T$ to $x_0$, which for sufficiently small diffusion steps can be approximated by Gaussian transitions:

$$q(x_{t-1}|x_t, y) \approx \mathcal{N}\left(x_{t-1}; \mu(x_t, t, y), \Sigma(x_t, t, y)\right), \quad (4)$$

where $y$ reflects the prior conditional information. A neural network $p_\theta$ can be trained to learn the noise-approximating inverse distribution by yielding approximations for $\mu_\theta$ and $\Sigma_\theta$ in Eq. (4) [42].

$$p_\theta(x_{t-1}|x_t, y) = \mathcal{N}\left(x_{t-1}; \mu_\theta(x_t, t, y), \Sigma_\theta(x_t, t, y)\right), \quad (5)$$

where the noisy image at each step, the model parameters, and the time step are denoted by $x_t$, $\theta$, and $t$, respectively [42]. And the learnable mean and covariance are given by $\mu_\theta(x_t, t, y)$, and $\Sigma_\theta(x_t, t, y)$. A commonly used parametrization excludes $\Sigma_\theta$ parameters to prioritize the focus on $\mu_\theta$:

$$\mu_\theta(x_t, t, y) = \frac{1}{\sqrt{\alpha_t}}\left(x_t - \frac{\beta_t}{\sqrt{1-\bar{\alpha}_t}}\varepsilon_\theta(x_t, t, y)\right) \quad (6)$$

The added noise $\varepsilon$ can be estimated by minimizing the L2 loss function, penalizing the error between the actual and predicted noise distributions:

$$L_{l2} = \mathbb{E}_{t, x_0, \varepsilon}\left[\left\|\varepsilon - \varepsilon_\theta\left(\sqrt{\bar{\alpha}_t}x_0 + \sqrt{1-\bar{\alpha}_t}\varepsilon, t, y\right)\right\|_2^2\right], \quad (7)$$

where $\mathbb{E}$ represents the expectation value.
To ensure that the inverse Markov chain closely approximates the forward process, it is essential to continuously adjust $\theta$ so that the joint distribution of the inverse process gradually aligns with that of the forward process. To minimize the discrepancy between $p_\theta(x_{t-1}|x_t, y)$ and the actual posterior probability during the forward process at each time step $t$, the following training objective, the negative ELBO with the parameter-independent terminal-prior term omitted, can be minimized:

$$L_{vlb} = \mathbb{E}_q\left[-log\, p_\theta(x_0|x_1, y) + \sum_{t>1}^{T} KL\left(q(x_{t-1}|x_t, x_0)||p_\theta(x_{t-1}|x_t, y)\right)\right], \quad (8)$$

where $KL$ represents the Kullback-Leibler divergence function. The expectation is taken over the training pairs and forward-diffusion samples. Thus, the overall generation loss function can be given by:

$$L_G = L_{l2} + \lambda_{vlb} \cdot L_{vlb}. \quad (9)$$

In practice, Eq. (9) is often simplified to $L_G = L_{l2}$, where the VLB term is omitted (i.e., $\lambda_{vlb} = 0$). This simplification has been shown to support high-fidelity image generation without the need to explicitly model the full data likelihood [43], [44].
Reverse diffusion steps are carried out during inference, beginning with a random sample $x_T \sim \mathcal{N}(0, I)$. For each step t, Eq. (6) is used to calculate $\mu$ based on the network estimate $\varepsilon_\theta$, whereas Eq. (5) is used to sample $x_{t-1}$.

### E. SCOD-GAN Architecture

We introduce a diffusion GAN model, SCOD-GAN, to efficiently perform the image translation task between the source and target OCTs. For fast and precise reverse diffusion sampling, SCOD-GAN benefits from a diffusive structure set up with a conditional adversarial training approach in two steps. The architecture, learning strategy, and inference process of the model are described below.

*1) Network's Architecture:* The model, as shown in Fig. 1, is designed for the sequential transformation of OCT images through two main stages: the transfer from PreOCT images $x_0^A$ (domain A) to PostMask images $x_0^B$ (intermediate domain B), and subsequently from PostMask to PostOCT images $x_0^C$ (domain C). The generator $G_{A\to B}$ maps PreOCT images to PostMask images, while the generator $G_{B\to C}$ maps PostMask images to PostOCT images. Both stages incorporate diffusion processes, where noise is added to intermediate representations to form noisy images, and reverse diffusion is performed conditionally using the respective generators to reconstruct the target images. Discriminators $D_B$ and $D_C$ are employed to differentiate between real and generated images in their domains. The model's training involves adversarial learning, where generators are optimized to minimize prediction errors and fool the discriminators, which are simultaneously trained to accurately distinguish between real images and generated ones.

The proposed two-stage architecture is designed to explicitly decouple macroscopic retinal morphometry from fine-grained OCT texture synthesis. First, Stage 1 predicts an intermediate post-treatment retinal mask, establishing a plausible global geometry and coarse spatial guidance. Conditioned on this anatomical prior, the Stage 2 generator then refines regional boundaries, resolves subtle fluid pockets, and synthesizes realistic retinal layers alongside their optical backscattering properties. This hierarchical division of labor spares a single generator from having to learn large-scale structural deformations and complex, tissue-specific OCT distributions simultaneously.
These two main stages are described as follows.

#### Step 1: Transfer from PreOCT to PostMask

*-Forward Diffusion from PreOCT to Intermediate Representation B:* Gaussian noise is added to the PostMask image $x_0^B$ to produce a noisy version $x_t^B$ at time step $t$ using the forward transition probability of:

$$q(x_t^B|x_{t-1}^B) = \mathcal{N}(x_t^B; \sqrt{1-\beta_t}\, x_{t-1}^B, \beta_t I). \quad (10)$$

*-Reverse Diffusion with Condition:* The neural network $G_{A\to B}$ is used to predict the $PostMask$ image from the noisy $PostMask$ image, conditioned on $PreOCT$ image presented by $y_B = x_0^A$. This can be expressed by:

$$x_0^B \sim G_{A\to B}(x_t^B, y_B = x_0^A, t). \quad (11)$$

*-Discriminator for PostMask:* The discriminator $D_B$ is used to differentiate between real PostMask images and those generated by $G_{A\to B}$.

#### Step 2: Transfer from PostMask to PostOCT

*-Forward Diffusion from Intermediate Representation B to PostOCT:* Noise is added to the $PostOCT$ image $x_0^C$ to produce a noisy version $x_t^C$ at time step $t$:

$$q(x_t^C|x_{t-1}^C) = \mathcal{N}(x_t^C; \sqrt{1-\beta_t}\, x_{t-1}^C, \beta_t I). \quad (12)$$

*-Reverse Diffusion with Condition:* The neural network $G_{B\to C}$ is used to predict the $PostOCT$ image from the noisy $PostOCT$

image, conditioned on $y_C = (x_0^A, \tilde{x}_0^B)$ where $\tilde{x}_0^B = G_{A\to B}(x_t^B, y_B = x_0^A, t)$. The reverse process is expressed as:

$$x_0^C \sim G_{B\to C}(x_t^C, y_c = (x_0^A, \tilde{x}_0^B), t). \quad (13)$$

*-Discriminator for PostOCT:* The discriminator $D_C$ is used to differentiate between real PostOCT images and those generated by $G_{B\to C}$.

*2) Supervised Learning Scheme with Adversarial Training:* To train the generators using a supervised end-to-end learning scheme, real PostMask and PostOCT images are used as ground truth. Generative loss functions are defined to measure the difference between the generated images and the real images in each domain.

**- $G_{A\to B}$ and $D_B$ Losses:** By utilizing loss Eq. (14), the generator $G_{A\to B}$ is optimized to reduce the discrepancy between the predicted PostMask image and the actual PostMask image, while also deceiving the discriminator $D_B$.

$$L_{G_{A\to B}} = \mathbb{E}_{t,q(x_t^B|x_0^B),y_B}\Big[BCE\left(x_0^B, G_{A\to B}(x_t^B, y_B = x_0^A, t)\right) + \left(1 - Dice\left(x_0^B, G_{A\to B}(x_t^B, y_B = x_0^A, t)\right)\right)\Big] + \mathbb{E}_{t,x_t^B}\Big[\log\left(1 - D_B(G_{A\to B}(x_t^B, y_B = x_0^A, t))\right)\Big], \quad (14)$$

where $BCE$ and $Dice$ refer to the Binary Cross-Entropy and Dice coefficient functions, respectively. Through Eq. (15), the discriminator $D_B$ simultaneously learns to differentiate between actual and synthetic PostMask images.

$$L_{D_B} = -\mathbb{E}_{x_0^B}[log\, D_B(x_0^B)] - \mathbb{E}_{t,x_t^B}[log(1 - D_B(G_{A\to B}(x_t^B, y_B = x_0^A, t)))]. \quad (15)$$

**- $G_{B\to C}$ and $D_C$ Losses:** The generator $G_{B\to C}$ is trained to minimize the error between the predicted PostOCT and the real PostOCT images, while simultaneously attempting to deceive the discriminator $D_C$. The discriminator $D_C$ is optimized to distinguish between real and generated PostOCT images:

$$L_{G_{B\to C}} = \mathbb{E}_{t,q(x_t^C|x_0^C),y_C}\left[\|x_0^C - G_{B\to C}(x_t^C, y_C = (x_0^A, \tilde{x}_0^B), t)\|_1\right] + \mathbb{E}_{t,x_t^C}\left[log\left(1 - D_C(G_{B\to C}(x_t^C, y_C = (x_0^A, \tilde{x}_0^B), t))\right)\right], \quad (16)$$

$$L_{D_C} = -\mathbb{E}_{x_0^C}[log\, D_C(x_0^C)] - \mathbb{E}_{t,x_t^C}\left[log\left(1 - D_C(G_{B\to C}(x_t^C, y_C = (x_0^A, \tilde{x}_0^B), t))\right)\right]. \quad (17)$$

The asymmetric loss design addresses stage-specific requirements. Stage 1 combines BCE and Dice losses with an adversarial objective to mitigate foreground–background class imbalance, improve boundary localization, and penalize anatomically implausible or disconnected retinal contours, thereby providing a reliable post-treatment retinal thickness prior. Conversely, Stage 2 combines an L1 reconstruction loss with an adversarial objective to synthesize continuous post-treatment OCT B-scans, promoting pixel-level intensity fidelity while preserving sharp retinal layer transitions and fine pathological microstructures, including fluid compartments and hyper-reflective foci, without the excessive smoothing often associated with L2 loss.

**- *Total Loss:*** The overall training objectives of the generators and discriminators are defined as weighted sums of the individual loss components:

$$L_G = \lambda_{mask}\cdot L_{G_{A\to B}} + \lambda_{image}\cdot L_{G_{B\to C}}, \quad L_D = \lambda_{mask}\cdot L_{D_B} + \lambda_{image}\cdot L_{D_C} \quad (18)$$

Each loss term is modulated by a task-specific weighting factor, which is tuned to balance competing objectives including post-total retina prediction accuracy and anatomical fidelity. In practice, the generators and discriminators are updated in an alternating fashion, each minimizing its own objective in Eqs. (14)-(17), with Eq. (18) specifying the task-specific weighting of the two stages.

The internal architecture of the UNet-based image generators used in this study involves a set of ResNet and Attention blocks, similar to NCSN++ [45]. Every block includes a conv-layer after two residual subblocks. The conv-layer doubles the channel dimension of every other block and lowers feature map resolution by half during the encoding process. For decoding, the conv-layer lowers the channel dimensions by half and increases the resolution for every other block. Temporal embedding of residual subblocks is obtained by passing a two-layer multi-layer perceptron (MLP) via a 32-dimensional sinusoidal position encoding [45]. Additionally, a 256-dimensional random latent is generated by a three-layer MLP and applied via adaptive normalization to modify feature maps [46]. Discriminators also utilize six blocks with two conv-layers, pooling, and the inclusion of temporal embedding to each block's feature maps.

*3) Inference Procedure:* During inference, the objective is to synthesize a PostOCT image from a given PreOCT image, utilizing only the generators within the model. For instance, to perform the mapping from PreOCT to PostMask (A→B), the generator $G_{A\to B}(x_t^B, y_B = x_0^A, t)$ is employed, where $x_t^B$ represents the PostMask sample of domain B at time step $t$, and $y_B$ is the acquired PreOCT image of source domain A used as a condition. To perform the subsequent mapping from PostMask to PostOCT (B→C), the generator $G_{B\to C}(x_t^C, y_C = (x_0^A, \tilde{x}_0^B), t)$ is utilized, where $x_t^C$ denotes the PostOCT-image sample of the target domain $C$ at time step $t$, and $y_C$ includes both the acquired PreOCT image of domain A and the synthesized PostMask of domain B as conditions. Inference begins at time step T with Gaussian noise samples $x_T^B$ and $x_T^C$ drawn from $\mathcal{N}(0, I)$. The noisy target-image samples generated at the end of each reverse diffusion step are used as the input target-image samples in the subsequent steps. A total of T reverse diffusion steps is performed to obtain $\tilde{x}_0^C$ at time step 0, which is the final synthetic PostOCT image.

*4) Response Prediction:* To determine individual therapeutic efficacy following SCOD-GAN OCT synthesis, cases are categorized as responsive or non-responsive based on baseline-to-post-treatment CST dynamics. Specifically, automated segmentation masks of both baseline and synthesized post-treatment B-scans are extracted using an established UNet++ model, thereby enabling fully automated, clinician-independent inference. A case is assigned to the responsive cohort if the resulting CST reduction exceeds a defined clinical threshold. Ground-truth response labels were independently derived from true post-treatment 3D-OCT volumes under ophthalmologist adjudication and reserved strictly for model validation.

## IV. Experimental Study And Results

### A. DME Datasets

In this study, the Duke-DME dataset was initially used to train and evaluate the proposed model against several baselines. To assess the model's generalizability and reduce dataset-specific bias, external validation was performed using the independent Isfahan-DME dataset, providing a more realistic evaluation of performance in real-world clinical scenarios.

## B. Evaluation Metrics

To comprehensively evaluate the performance of the models, we used a variety of established metrics, including the well-known MAE [47], peak signal-to-noise ratio (PSNR) [48], structural similarity index measure (SSIM) [49], and Fréchet inception distance (FID) [50]. We also used the sliced Wasserstein distance (SWD) to assess the similarity between image distributions [51], the inception score (IS) to evaluate both the quality and diversity of the generated images based on their recognizability by a trained classifier [52], and kernel inception distance (KID) as a non-parametric alternative to FID, providing an unbiased estimation of distribution difference [53]. We further evaluated our model's classification performance in predicting patient response to the treatment, categorized as responsive (CST percentage change > threshold) or non-responsive (CST percentage change ≤ threshold). Two thresholds (10% and 20% CST reduction) were investigated as operational definitions of OCT-derived anatomical treatment response [54], [55]. The percentage change in CST was calculated as follows:

$$CST\ Change\ (\%) = \frac{V_1 - V_2}{V_1} \times 100, \tag{19}$$

where $V_1$ and $V_2$ are 1-*mm* diameter central retinal thickness volumes in *mm³* obtained from pre- and post-treatment 3D-OCT scans. For the fixed 1-mm subfield area, the percentage change in this volume equals the percentage change in mean CST. Classification performance was evaluated using balanced accuracy (Acc), precision (Pr), sensitivity (Se), and F1-score metrics.

## C. Implementation Details

Our experiments were implemented using PyTorch version 1.12 on a cluster of two TITAN GPUs. Each weight layer includes spectrum normalization and LeakyReLU activation.

To quantitatively evaluate the performance of the proposed model and the baselines, five-fold cross-validation (CV) was performed using random subject-level partitioning without stratification. To prevent data leakage, all OCT scans from the same subject were assigned to the same fold, ensuring that no B-scans from the same subject appeared in both the training and testing sets during any iteration. For optimizing the generative models on the Duke-DME dataset, the central 49 B-scans from each training OCT volume (6,223 total B-scans across 127 patients) were used to ensure a consistent region of analysis, minimizing the impact of artifacts that may occur in peripheral scans.

The hyperparameters used for the 5-fold CV method were 100 epochs, a learning rate of 1.5e-4, 4 diffusion steps, and $\lambda_{mask} = \lambda_{image} = 1$. The diffusion step hyperparameter was optimized in terms of visual fidelity—specifically using the SSIM metric—via grid search over the range [2, 4, 10, 50, 1000] on the first training subset of the Duke-DME dataset. A similar strategy was applied to optimize $\lambda_{mask}$ and $\lambda_{image}$, where grid search was conducted over the range [0.01, 0.25, 0.5, 1.0, 1.5], using the F1-score as the evaluation metric. Additionally, the noise variance was bounded by lower and higher limits of to $\beta_{min} = 0.1$, $\beta_{max} = 20$, respectively.

## D. Competing Methods and State-of-the-art Analysis

We evaluated the performance of our proposed model by comparing it to several well-established and recently developed baselines that have significantly influenced the fields of natural and medical image generation and translation. For baseline analysis, we considered prominent methods, including Pix2Pix [21], CycleGAN [22], UNIT [23], MUNIT [24], BicycleGAN [25], and Pix2PixHD [26]. To assess state-of-the-art (SOTA) performance, we included models such as HengSRC [27], Fast-DDPM [29], SGD [30], I2I-Mamba [31], selfRDB [32], FGDM [33], PatchGCL [34], and StegoGAN [35]. These baseline methods provide a robust foundation for assessing the performance of our proposed OCT image translation model. The experimental performance results for OCT-to-OCT image translation and class prediction tasks are detailed in Tables II and III, respectively. Fig. 3 illustrates test sample images, demonstrating the SCOD-GAN framework's effectiveness in generating post-treatment OCT B-scans and corresponding binary masks.

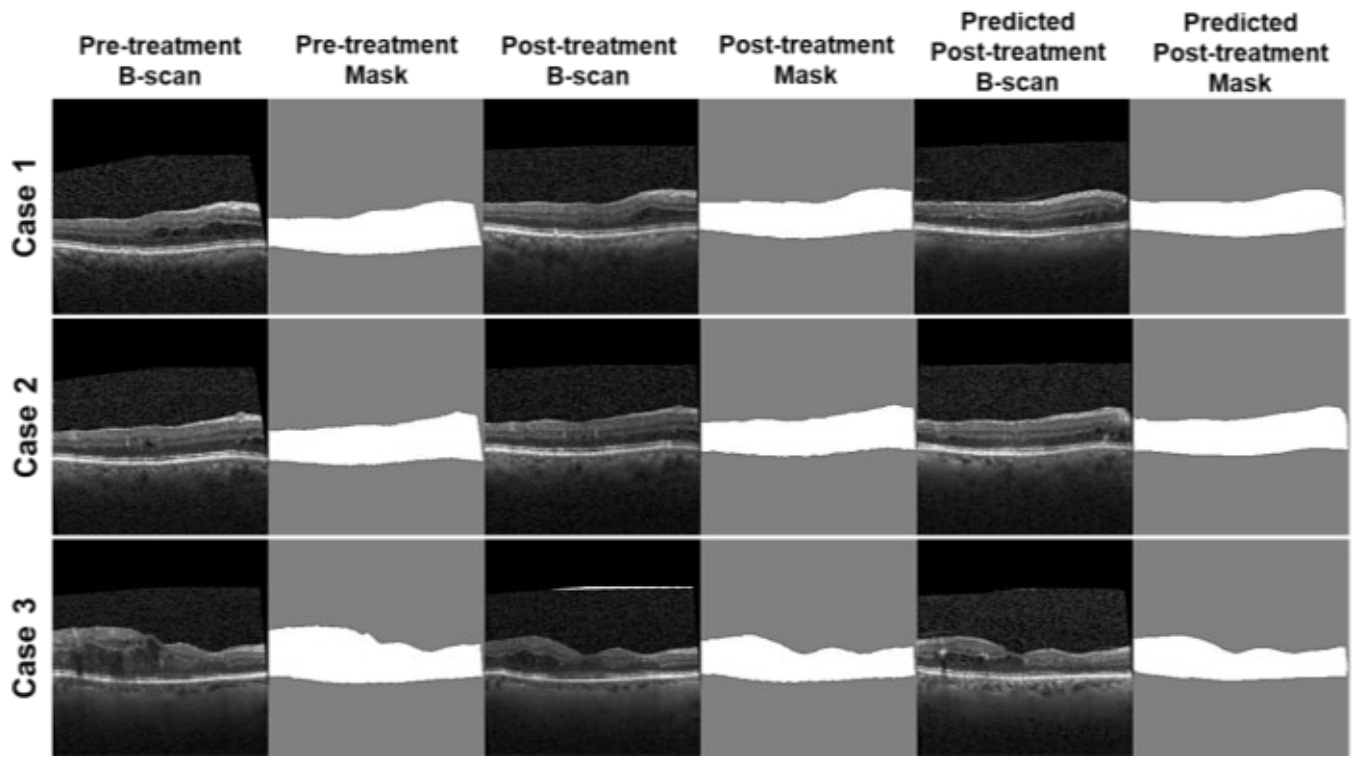

Fig. 3. Test sample visualization to determine the qualitative performance of the SCOD-GAN framework in generating post-treatment OCT B-scans and their corresponding binary masks in the Duke-DME dataset.

## E. Ablation Study Design

To evaluate the contributions of the SCOD-GAN components, a comprehensive ablation study was conducted.

*1) Effectiveness of Structural Components:* Four alternative configurations were systematically evaluated against our full model using the Duke-DME dataset. First, the efficacy of the two-stage approach was verified by comparing the performance of the full model against a single-stage baseline (*w/o Two-Stage Process*) designed to directly and conditionally map pre- to post-treatment images. Second, the impact of adversarial training was isolated through a variant (*w/o Adversarial Loss*) engineered to be trained exclusively with the pixel-wise reconstruction loss, thereby removing both discriminators. Third, the importance of intermediate supervision was investigated by ablating the mask discriminator, $D_B$ (*w/o Mask Discriminator*), which isolated the contribution of adversarial supervision on the intermediate mask, while the direct BCE-Dice mask supervision remained in place. Finally, the contribution of the diffusion-based architecture was assessed by benchmarking against a model (*w/o Diffusion Modules*) in which the core diffusive modules were replaced with standard UNet generators and PatchGAN discriminators [21]. The performance of each configuration was quantitatively assessed, and the ablation results are reported in Table IV.

TABLE II
PERFORMANCE COMPARISON OF THE SCOD-GAN MODEL WITH BASELINE AND SOTA FRAMEWORKS IN THE RETINAL OCT-TO-OCT IMAGE TRANSLATION TASK. BEST MEAN RESULTS ARE HIGHLIGHTED IN BOLD.

| Data | Model | Performance | | | | | | |
|---|---|---|---|---|---|---|---|---|
| | | MAE (↓) $\times 10^{-3}$ | PSNR (↑) | SSIM (↑) $\times 10^{-2}$ | FID (↓) | SWD (↓) $\times 10^{-4}$ | IS (↑) | KID (↓) $\times 10^{-4}$ |
| Duke-DME | Pix2Pix [21] | 103 | 20.53 | 78.35 | 7.541 | 39 | 1.0132 | 241 |
| | CycleGAN [22] | 98 | 20.70 | 78.46 | 8.622 | 42 | 1.0131 | 260 |
| | UNIT [23] | 94 | 20.85 | 78.53 | 8.552 | 41 | 1.0134 | 258 |
| | MUNIT [24] | 92 | 20.94 | 78.77 | 7.918 | 39 | 1.0133 | 232 |
| | BicycleGAN[25] | 91 | 20.96 | 80.12 | 6.983 | 34 | 1.0145 | 209 |
| | Pix2PixHD [26] | 89 | 20.97 | 80.19 | 6.725 | 30 | 1.0157 | 171 |
| | Heng-SRC [27] | 72 | 22.27 | 81.27 | 1.918 | **16** | 1.0199 | 89 |
| | Fast-DDPM [29] | 86 | 21.23 | 80.69 | 7.781 | 39 | 1.0147 | 202 |
| | SGD [30] | 83 | 21.74 | 80.76 | 6.746 | 30 | 1.0164 | 178 |
| | I2I-mamba [31] | 85 | 20.91 | 80.29 | 6.517 | 23 | 1.0238 | 169 |
| | SelfRDB [32] | 86 | 20.96 | 80.21 | 6.607 | 26 | 1.0189 | 198 |
| | FGDM [33] | 79 | 21.66 | 80.71 | 2.863 | 19 | 1.0163 | 125 |
| | PatchGCL [34] | 77 | 21.92 | 80.88 | 1.899 | 17 | 1.0210 | 66 |
| | StegoGAN [35] | 76 | 22.08 | 80.96 | 1.719 | 17 | 1.0246 | 51 |
| | SCOD-GAN (Ours) | **51** | **23.98** | **83.41** | **1.655** | 17 | **1.0268** | **35** |
| Isfahan-DME (External Validation) | Pix2Pix [21] | 114 | 19.15 | 72.82 | 10.875 | 61 | 1.0140 | 359 |
| | CycleGAN [22] | 96 | 19.67 | 73.70 | 10.364 | 58 | 1.0148 | 342 |
| | UNIT [23] | 95 | 19.25 | 73.56 | 10.276 | 59 | 1.0159 | 328 |
| | MUNIT [24] | 95 | 19.56 | 74.32 | 9.411 | 52 | 1.0163 | 295 |
| | BicycleGAN[25] | 92 | 19.71 | 76.50 | 6.956 | 44 | 1.0167 | 234 |
| | Pix2PixHD [26] | 90 | 19.98 | 77.30 | 7.021 | 45 | 1.0158 | 206 |
| | Heng-SRC [27] | 78 | 20.36 | 77.93 | **1.949** | 29 | 1.0171 | 111 |
| | Fast-DDPM [29] | 90 | 19.42 | 77.47 | 8.147 | 43 | 1.0147 | 252 |
| | SGD [30] | 89 | 19.96 | 77.38 | 6.926 | 39 | 1.0164 | 239 |
| | I2I-mamba [31] | 82 | 19.85 | 77.67 | 5.390 | 34 | 1.0182 | 196 |
| | SelfRDB [32] | 83 | 20.25 | 77.45 | 5.972 | 33 | 1.0166 | 208 |
| | FGDM [33] | 82 | 20.09 | 75.74 | 5.404 | 31 | 1.0175 | 197 |
| | PatchGCL [34] | 80 | 20.36 | 77.79 | 2.031 | **27** | 1.0191 | 107 |
| | StegoGAN [35] | **73** | **20.80** | 78.68 | 2.680 | 30 | 1.0189 | 122 |
| | SCOD-GAN (Ours) | 76 | **20.80** | **79.09** | 1.991 | 28 | **1.0217** | **97** |

** **Experimental Setup:*** The models were first optimized and evaluated on the Duke-DME dataset, with performance averages calculated using the 5-fold cross-validation method. For external validation on the Isfahan-DME dataset, the results represent the average performance of the optimized models across each iteration of the cross-validation process.
*** **Inference time (per image) comparison on a standard PC*** (CPU: AMD Ryzen 7 3700X 8-Core Processor, RAM: 32GB, OS: Ubuntu 22.04 LTS, GPU: Nvidia GeForce RTX 3060 12GB): Heng-SRC, I2I-mamba, and PatchGCL are the fastest, with inference times of 0.004s, whereas SGD and Fast-DDPM are the slowest, taking 93.57s and 11.55s, respectively. FGDM and SCOD-GAN have moderate inference times of 0.439s and 0.427s, respectively.

TABLE III
ANTI-VEGF TREATMENT RESPONSE PREDICTION PERFORMANCE RESULTS OF THE BEST-PERFORMING TRANSLATION MODELS. BEST MEAN RESULTS ARE HIGHLIGHTED IN BOLD.

| Data | Model | Response Prediction Performance according to Volume Thickness Thr (%)=[10, 20] | | | |
|---|---|---|---|---|---|
| | | Acc (%) | Pr (%) | Se (%) | F1 (%) |
| Duke-DME | Heng-SRC [27] | 69.42, 74.11 | 55.84, 79.63 | 66.67, 76.47 | 60.78, 78.02 |
| | Fast-DDPM [29] | 68.96, 67.35 | 56.39, 76.97 | 64.08, 67.34 | 59.98, 71.83 |
| | SGD [30] | 71.69, 68.06 | 60.11, 78.57 | 66.76, 64.71 | 63.26, 70.96 |
| | I2I-mamba [31] | 63.43, 66.94 | 57.14, 81.80 | 44.45, 52.94 | 50.01, 64.28 |
| | SelfRDB [32] | 67.65, 69.36 | 52.87, 76.94 | 66.67, 75.23 | 58.97, 76.08 |
| | FGDM [33] | 70.75, 72.29 | 54.01, 77.27 | 75.36, 85.12 | 62.92, 81.01 |
| | PatchGCL [34] | 79.32, 74.68 | 69.87, 78.55 | 75.91, 88.46 | 72.76, 83.21 |
| | StegoGAN [35] | 80.22, 84.39 | 83.90, 90.16 | 67.24, 83.55 | 74.66, 86.73 |
| | SCOD-GAN (Ours) | **85.69, 92.98** | **89.76, 95.34** | **75.97, 93.35** | **82.29, 94.32** |
| Isfahan-DME (External Validation) | Heng-SRC [27] | 69.84, 89.68 | 56.78, 81.59 | 65.45, 84.28 | 60.81, 82.90 |
| | Fast-DDPM [29] | 68.53, 78.66 | 55.96, 79.60 | 62.50, 61.37 | 59.05, 69.31 |
| | SGD [30] | 71.89, 86.78 | 61.94, 78.46 | 64.18, 79.14 | 63.04, 78.80 |
| | I2I-mamba [31] | 62.28, 76.12 | 56.25, 72.61 | 41.09, 57.85 | 47.49, 64.41 |
| | SelfRDB [32] | 70.82, 89.45 | 57.26, 77.09 | 67.83, 85.42 | 62.09, 81.04 |
| | FGDM [33] | 73.04, 91.31 | 59.51, 79.87 | 71.11, 88.35 | 64.79, 83.90 |
| | PatchGCL [34] | 79.78, 91.92 | 71.84, 81.75 | 74.72, 88.95 | 73.25, 85.21 |
| | StegoGAN [35] | 79.70, 91.51 | 81.96, 87.38 | 67.04, 86.23 | 73.81, 86.80 |
| | SCOD-GAN (Ours) | **85.17, 95.25** | **88.61, 94.09** | **75.36, 91.98** | **81.46, 93.02** |

TABLE IV
QUANTITATIVE RESULTS OF THE ABLATION STUDY ON DUKE-DME DATASET, EVALUATING THE CONTRIBUTION OF EACH KEY COMPONENT IN THE SCOD-GAN. BEST MEAN RESULTS ARE HIGHLIGHTED IN BOLD.

| Model Configuration | Performance | | | | | |
|---|---|---|---|---|---|---|
| | OCT Image Translation | | | | PostMask Generation | Response Prediction Thr=[10, 20] |
| | MAE (↓) $\times 10^{-3}$ | PSNR (↑) | SSIM (↑) $\times 10^{-2}$ | FID (↓) | Dice (↑) | F1 (↑) (%) |
| Full Model: SCOD-GAN | **51** | **23.98** | **83.41** | **1.655** | **0.9457** | **82.29, 94.32** |
| w/o Two-Stage Process | 63 | 22.24 | 81.01 | 2.516 | — | 77.49,89.16 |
| w/o Adversarial Loss | 71 | 21.59 | 80.98 | 3.117 | 0.9164 | 76.94,88.24 |
| w/o Mask Discriminator | 56 | 23.11 | 82.15 | 1.987 | 0.9288 | 79.68,91.86 |
| w/o Diffusion Models | 75 | 21.33 | 80.47 | 3.254 | 0.8998 | 75.17,87.59 |

TABLE V
COMPARATIVE PERFORMANCE AND INTERPRETABILITY ANALYSIS OF SINGLE-STAGE TRANSLATION AND TWO-STAGE SCOD-GAN ON THE DUKE-DME DATASET USING THE 5-FOLD CV METHOD.

| Analysis Dimension | Metric | Single-stage | Two-stage | Relative Gain / Association | P-value* |
|---|---|---|---|---|---|
| Global Translation Quality | MAE×$10^{-3}$ (↓) | 89±11 | 82±9 | 7.8% ↓ | <0.01 |
| | PSNR (dB) (↑) | 20.71±2.3 | 21.38±1.7 | +0.67 dB | 0.04 |
| | SSIM×$10^{-2}$ (↑) | 79.63±4.1 | 81.95±2.6 | +2.32% points | <0.01 |
| | FID (↓) | 2.556±0.9 | 2.379±0.4 | 6.9% ↓ | — |
| Why does it help? (Oracle Mask Conditioning) | MAE with predicted PostMask (↓) | — | 82±9 | Reference condition | — |
| | MAE with GT PostMask (↓) | — | 79±7 | 3.2% ↓ vs. predicted mask | 0.06 |
| Where do gains occur? (Spatial Decomposition) | Retinal MAE×$10^{-3}$ (↓) | 63±10 | 51±7 | 19.0% ↓ | <0.01 |
| | Non-retinal MAE×$10^{-3}$ (↓) | 91±14 | 88±12 | 3.3% ↓ | 0.12 |
| How does Stage 1 affect Stage 2? (Error Coupling) | Spearman correlation: *PostMask Error vs. PostOCT MAE* | — | ρ= 0.41 | Positive Association | 0.03 |
| | Spearman correlation: *PostMask Error vs. PostOCT SSIM* | — | ρ=−0.56 | Negative Association | <0.01 |
| Clinical Relevance | Central subfield volume error (mm$^3$) =$\lvert V_{GT}-V_{pred}\rvert$ (↓) | 0.0217 | 0.0190 | 0.0027 mm$^3$ ↓ (12.4%) | <0.01 |
| Image-to-Clinical Metric Coupling | Spearman correlation: *PostOCT MAE vs. Volume Error* | ρ=0.49 | ρ=0.52 | Positive Association | <0.01 |
| | Spearman correlation: *PostOCT SSIM vs. Volume Error* | ρ=−0.41 | ρ=−0.57 | Negative Association | <0.01 |

** **Notes:*** Statistical significance evaluated using *Wilcoxon signed-rank tests* for case-level metric differences and *Spearman rank correlation* for association analyses, p<0.05 considered statistically significant. Global Translation Quality metrics are computed over the full B-scan, in contrast to the retina-masked metrics reported in Tables II and IV.

*2) Comparative Interpretability Analysis of the Two-Stage Design:* To investigate the mechanistic basis and spatial distribution of performance gains achieved by our two-stage cascaded architecture over the single-stage baseline (*w/o Two-Stage Process*), we conducted a structured interpretability study on the Duke-DME dataset across five analytical dimensions (summarized in Table V): (i) Global Translation Quality, evaluated across the full B-scan using MAE, PSNR, SSIM, and FID; (ii) Oracle Mask Conditioning (*"Why does it help?"*), assessed by conditioning the Stage 2 generator ($G_{B\to C}$) on ground-truth post-treatment masks ($x_0^{B,\mathrm{GT}}$) versus Stage 1 predicted masks ($\tilde{x}_0^B$) to isolate the causal contribution of explicit boundary guidance; (iii) Spatial Error Decomposition (*"Where do gains occur?"*), performed by partitioning pixel-wise reconstruction error into the clinically vital neurosensory retinal compartment (Retinal MAE, spanning from the NFL to the RPE) versus background regions (Non-retinal MAE); (iv)

Inter-Stage Error Coupling (*"How does Stage 1 affect Stage 2?"*), quantified by computing the association between Stage 1 mask prediction error and Stage 2 synthesis metrics (PostOCT MAE and SSIM); and (v) Clinical Relevance and Metric Coupling, evaluated via the absolute central subfield volume error and its correlation with OCT image-level fidelity metrics.

*3) Impact of the Loss Coefficients on Model Performance:* To determine the optimal balance between the constituent objectives of the SCOD-GAN architecture, a systematic sensitivity analysis was performed. It was conducted on the Duke-DME dataset to determine optimal values for the loss coefficients $\lambda_{mask}$ and $\lambda_{image}$ (Eq. (18)). Each coefficient was independently varied across the set [0.01, 0.25, 0.5, 1.0, 1.5] to evaluate its impact on PostMask generation, PostOCT synthesis, and treatment response prediction performance, respectively. This analysis, detailed in Fig. 5, elucidates the model's sensitivity to these hyperparameters.

## V. Discussion

Predicting anti-VEGF treatment efficacy via post-treatment CST on retinal OCT is pivotal for tailoring first-line therapies in DME. Building on prior generative frameworks [19], [20], this study introduces SCOD-GAN, a source-conditional adversarial generative model integrated with thickness-aware conditional diffusion to achieve high-fidelity retinal OCT image translation and accurate treatment response prediction.

### A. Baseline Analysis for Pre-to-Post Treatment OCT Image Translation

The image translation performance of the proposed SCOD-GAN model was rigorously compared against several baseline and SOTA models across the datasets, as summarized in Table II. To focus our evaluation on the clinically relevant region, we analyzed the total retinal area, from the NFL to the RPE layers, in each B-scan using the specifically trained UNet++ segmentation model, which was validated by our expert ophthalmologist (A.D.). Subsequently, the image generation metrics were computed exclusively within this region to avoid the influence of unpredictable eye motion in non-retinal areas. The auxiliary UNet++ network was trained and cross-validated exclusively on SD-OCT volumes from the Duke-DME dataset, with the external Isfahan-DME cohort held strictly as an unseen test set. It achieved a mean Bscan-level Dice coefficient of 0.958±0.011 and 0.944±0.036 on the Duke- and Isfahan-DME datasets, respectively.

The SCOD-GAN model exhibited superior performance across both datasets, outperforming or matching the baselines in several critical evaluation metrics. On the Duke-DME dataset, SCOD-GAN achieved the lowest MAE ($51\pm7\times10^{-3}$) and the highest PSNR (23.98±0.51), demonstrating its ability to produce post-treatment images with high fidelity, promising accuracy, and clarity. Additionally, SCOD-GAN recorded the highest SSIM ($83.41\pm0.9\times10^{-2}$), indicating better preservation/translation of key structural details within the generated post-treatment images. The model also led in FID (1.655±0.14), IS (1.0268±0.003), and KID ($35\pm7\times10^{-4}$), with the lowest scores among all models, demonstrating that its generated images closely aligned with the real image distribution, thus ensuring high quality and realism.

Notably, SCOD-GAN outperformed advanced models, including SelfRDB, FGDM, PatchGCL, and StegoGAN. On the Isfahan-DME dataset, SCOD-GAN maintained acceptable performance, closely trailing the top-performing models. While Heng-SRC achieved a slightly lower FID (1.949±0.22 vs. 1.991±0.26), PatchGCL yielded a marginally lower SWD ($27\pm2\times10^{-4}$ vs. $28\pm3\times10^{-4}$), and StegoGAN attained a comparable PSNR (20.80±0.7 vs. 20.80±0.6) and a slightly lower MAE ($73\pm8\times10^{-3}$ vs. $76\pm5\times10^{-3}$), SCOD-GAN still surpassed other models in several areas, particularly in generating images that are not only accurate but also realistic and diverse. This is evidenced by its competitive SSIM, FID, and KID scores, demonstrating a strong balance of high quality across various evaluation metrics.

For the sake of completeness, the qualitative performance of our SCOD-GAN model is compared with the best-performing SOTA methods for pre-to-post OCT image translation in Fig. 4. The figure assesses the generative models using the datasets and encompasses both responsive and non-responsive samples. The results demonstrate that SCOD-GAN visually surpasses other models by generating translated images that are sharper, more accurately translate retinal anatomical structures, and exhibit greater similarity to the real post-treatment images for both responsive and non-responsive samples. Corresponding pixel-wise absolute error maps for SCOD-GAN and competing SOTA baselines are shown in Fig. S2 of the Supplementary Material.

### B. Performance Analysis for Response Prediction

The performance of the SCOD-GAN framework in predicting treatment response was evaluated and compared across the datasets. The results from Table III, detailing performance scores, reveal that the SCOD-GAN framework demonstrates promising performance in predicting treatment response based on thickness volume measure, with the highest Acc, Pr, Se, and F1-score at both 10% and 20% reduction thresholds. Moreover, external validation on the Isfahan-DME dataset reaffirmed the effectiveness of the model. Although performance declined during external validation compared to the Duke-DME dataset, highlighting potential cross-dataset domain shifts, SCOD-GAN maintained competitive performance, demonstrating its acceptable robustness. A quantitative cross-center analysis of demographic, clinical, technical, and feature-space distributional shifts is provided in Section 1 of the Supplementary Material.

To assess the extent to which minor UNet++ segmentation errors propagate to downstream clinical endpoints, we conducted a sensitivity analysis to evaluate the robustness of treatment-response classification to boundary perturbations. Specifically, controlled dilation and erosion of the NFL and RPE boundaries by up to ±3 pixels altered the assigned response category in fewer than 1.6% of cases in the Duke-DME dataset, demonstrating the robustness of the downstream classification to segmentation uncertainty.

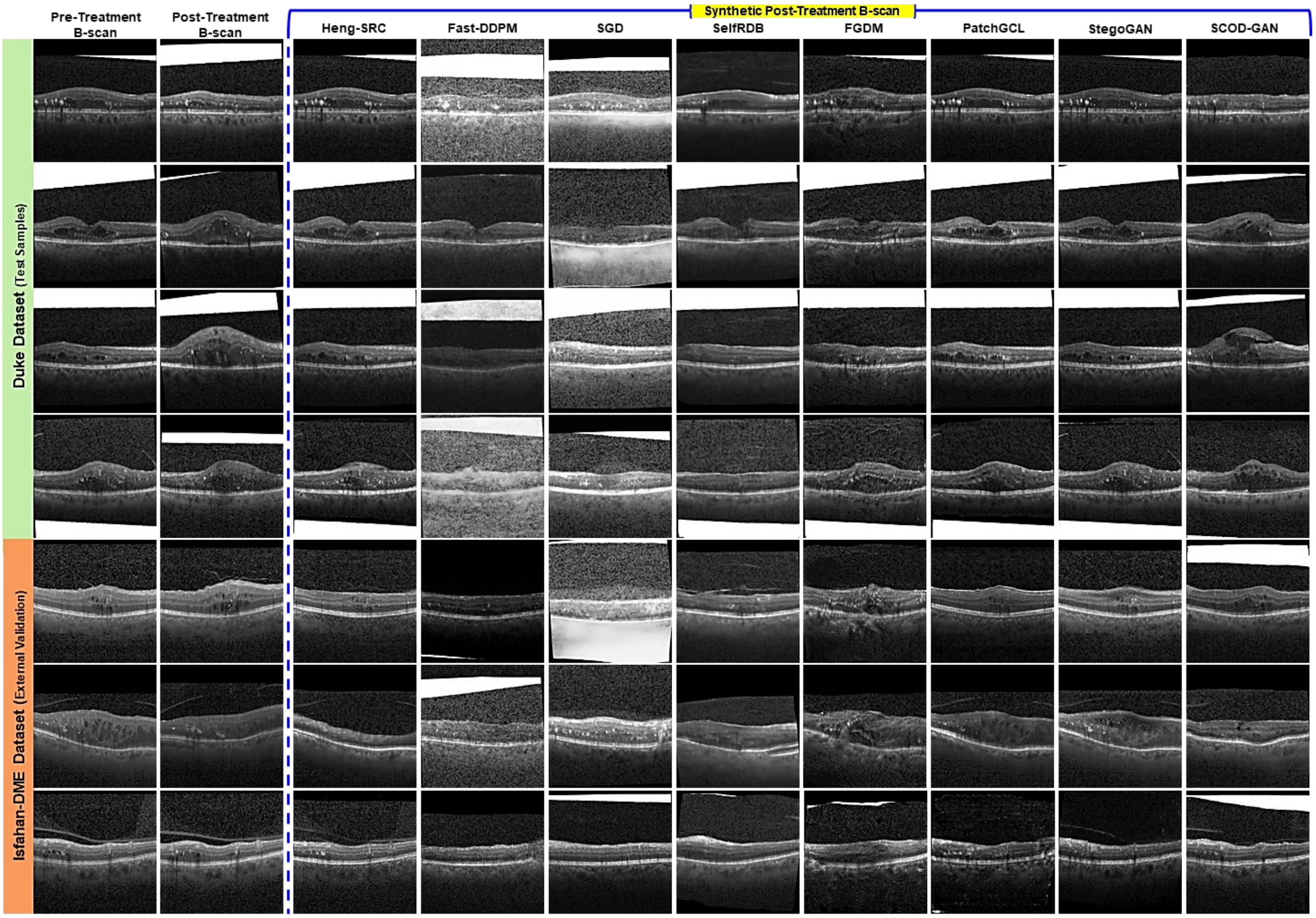

Fig. 4. Performance of the SCOD-GAN framework in generating post-treatment OCT B-scans, including both responsive and non-responsive samples, compared to competing methods. The proposed model distinctly outperforms similar frameworks.

### C. Ablation Study Analysis

*1) Network Structure Design:* The ablation results in Table IV demonstrate how each architectural component contributes to the overall performance of SCOD-GAN. Comparing the full model against its single-stage counterpart (*without the two-stage process*) highlights the value of decomposing the task into intermediate mask prediction and conditional synthesis.

Removing the first stage consistently degrades performance: MAE increases from $51\pm7\times10^{-3}$ to $63\pm10\times10^{-3}$ and SSIM drops from $83.41\pm0.9\times10^{-2}$ to $81.01\pm1.6\times10^{-2}$, with even sharper declines in FID (1.655±0.14 to 2.516±0.37) and F1-score (82.29±4.7%, 94.32±2.6%) to (77.49±5.3%, 89.16±4.9%). These findings confirm that direct, single-step OCT-to-OCT translation is overly complex, whereas our two-stage design better captures structural and pathological features to yield clinically relevant outputs.

Omitting the adversarial loss and relying solely on pixel-wise reconstruction further degrades performance, raising MAE to $71\pm10\times10^{-3}$ and FID to 3.117±0.68. While SSIM remains comparable ($80.98\pm1.8\times10^{-2}$), the F1-score falls to (76.94±6.6%, 88.24±5.8%) and the Dice coefficient drops from 0.9457±0.051 to 0.9164±0.072. Although pixel-wise losses enforce coarse structural alignment, they tend to produce over-smoothed images lacking high-frequency OCT textures; adversarial supervision is therefore essential to align the generator with the true OCT distribution and accurately predict pathological response regions. Similarly, ablating the mask discriminator causes a moderate performance drop—reducing the Dice score to 0.9288±0.063 and F1-score to (79.68±5.9%, 91.86±4.1%) which confirms that explicit intermediate supervision prevents visually plausible yet anatomically inaccurate PostOCT generations, thereby safeguarding downstream clinical predictions.

The most severe degradation occurs when replacing the diffusion modules with a conventional UNet generator and PatchGAN discriminator (*w/o diffusion*): MAE reaches $75\pm13\times10^{-3}$, PSNR drops to 21.33±0.72, SSIM falls to $80.47\pm1.9\times10^{-2}$, and FID peaks at 3.254±0.69, accompanied by the lowest Dice (0.8998±0.078) and F1-scores (75.17±6.6%, 87.59±5.9%). Unlike standard UNet translations, the iterative denoising inherent to diffusion models effectively captures complex variability and subtle pathological textures in OCT scans. Ultimately, SCOD-GAN's effectiveness stems not from a single module, but from the tight integration of its two-stage paradigm, diffusive generation, and multi-level adversarial supervision.

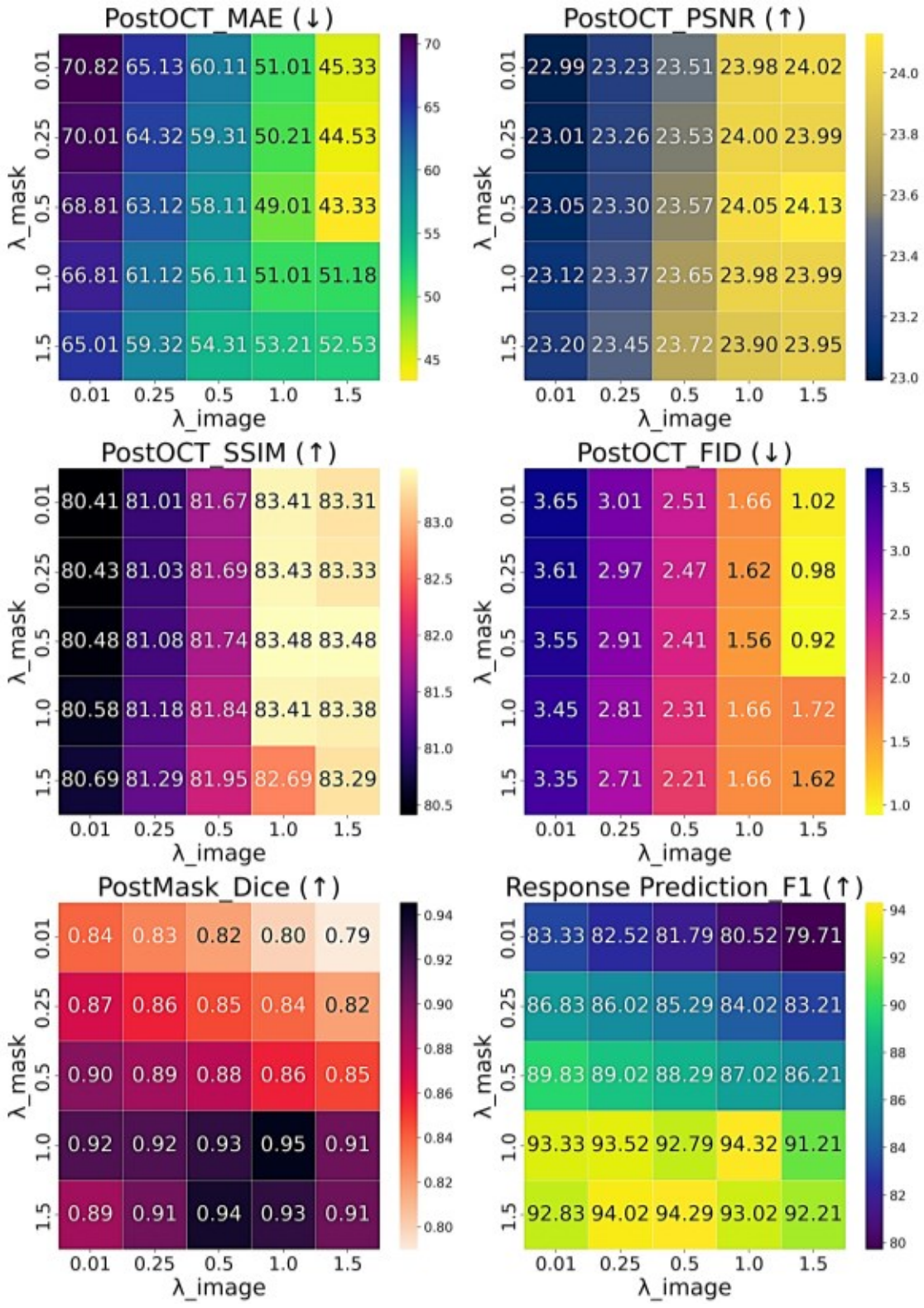

Fig. 5. Heatmap visualization of $\lambda_{mask}$ and $\lambda_{image}$ weights hyperparameter tuning for the proposed model. The plots show how adjusting $\lambda_{mask}$ and $\lambda_{image}$ weights affect image synthesis (PostOCT metrics), mask generation (PostMask_Dice), and final response prediction (F1), revealing the optimal configuration for peak performance.

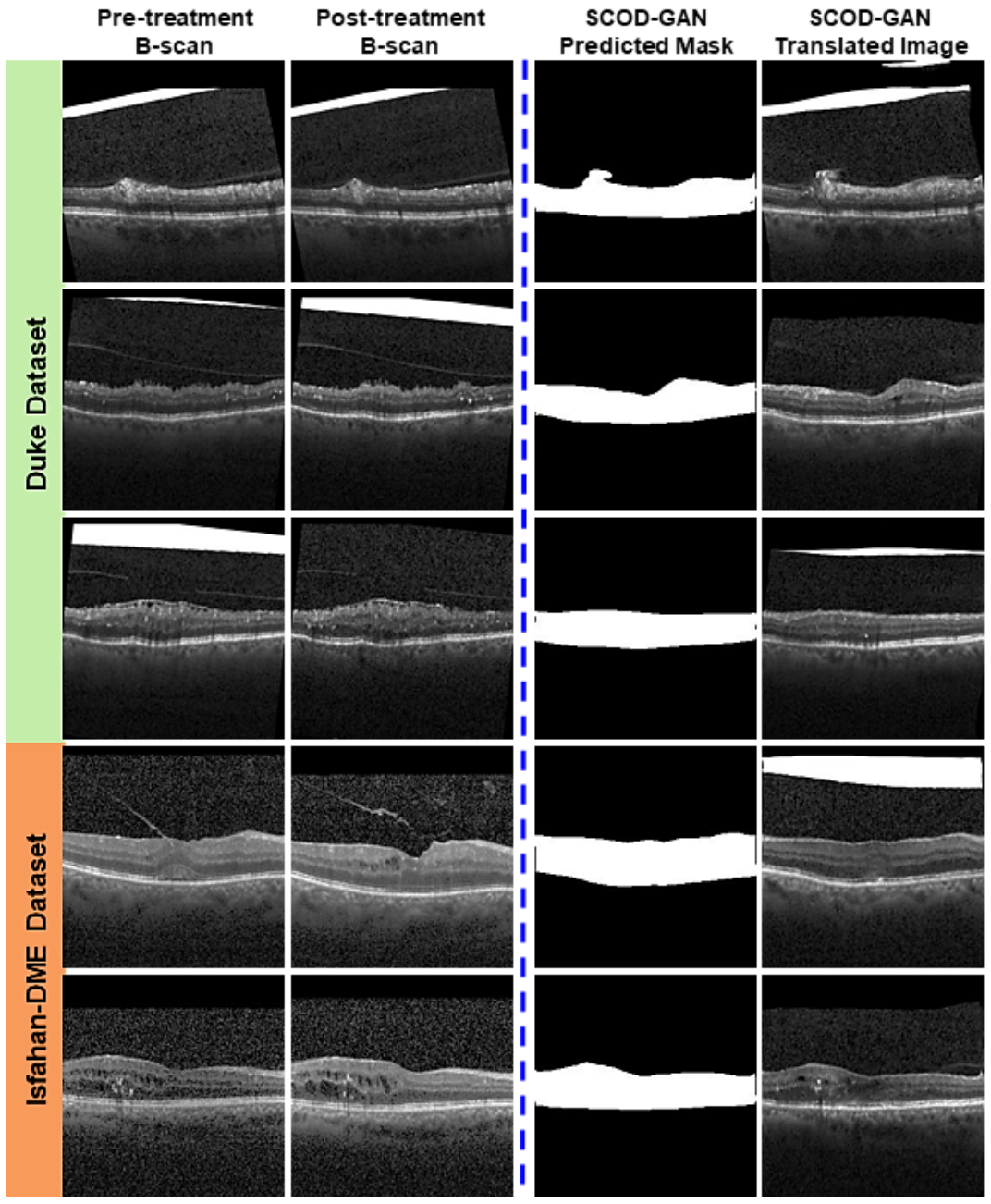

Fig. 6. Qualitative analysis of the proposed model's primary failure mode on borderline cases. This figure displays representative examples where SCOD-GAN failed to correctly classify the treatment response. Each row corresponds to a single case from the test sets. These failure cases are characterized by a marginal change in CST of approximately ±5%, representing challenging borderline scenarios where the model struggles with both PostOCT image synthesis and anti-VEGF treatment response prediction.

*2) Interpretability of Architectural Decoupling:* The comparative interpretability results in Table V clarify "*where*", "*why*", and "*how*" decoupling macroscopic structural morphometry (*Stage 1*) from microstructural texture synthesis (*Stage 2*) outperforms direct single-stage translation. Spatially, while the two-stage pipeline yields a modest, non-significant error change in noisy non-retinal background regions (3.3% reduction in Non-retinal MAE, p=0.12), it achieves a highly pronounced 19.0% error reduction within the neurosensory retinal layer (Retinal MAE reduced from $63 \pm 10 \times 10^{-3}$ to $51 \pm 7 \times 10^{-3}$, p<0.01), alongside a 7.8% reduction in global MAE (p<0.01) and a +2.32% point gain in SSIM (p<0.01). This demonstrates that the intermediate mask ($\tilde{x}_0^B$) functions as an explicit spatial prior that concentrates generative optimization within the entire retinal region. Mechanistically, conditioning Stage 2 on ground-truth masks ($x_0^{B,\mathrm{GT}}$) further reduces global MAE by 3.2% (p=0.06), while Stage 1 mask errors correlate significantly with Stage 2 PostOCT MAE (ρ=+0.41, p=0.03) and SSIM (ρ=-0.56, p<0.01), confirming that Stage 1 serves as a transparent morphological bottleneck. Clinically, this targeted structural guidance reduces the absolute central subfield volume error by 0.0027 $mm^3$ (12.4% reduction, from 0.0217±0.0104 to 0.0190±0.0077 $mm^3$, p<0.01) and tightens the coupling between image-level fidelity and clinical thickness accuracy (ρ=-0.57 vs. -0.41 for SSIM vs. Volume Error, and ρ=+0.52 vs. 0.49 for MAE vs. Volume Error; p<0.01), ensuring that higher perceptual and structural synthesis quality translates into reliable anti-VEGF response prediction.

*3) The Loss Coefficients Analysis:* The experimental study on the $\lambda_{\text{mask}}$ and $\lambda_{image}$ values, reveals a critical trade-off between generative fidelity and treatment response prediction within the SCOD-GAN framework. As visualized in Fig. 5, heavily weighting the PostOCT image synthesis loss ($\lambda_{image}$) predictably enhances image quality metrics (MAE, PSNR, SSIM, FID), yet this prioritization does not yield optimal predictive performance. The ultimate predictive power, measured by the F1-score, is instead contingent upon a balanced integration of both generation losses. Our findings demonstrate that the weighting of $[\lambda_{mask}, \lambda_{image}] = [1,1]$ establishes an optimal balance. At this point, the feature representations learned for both tasks become mutually reinforcing, leading to a peak F1-score of 94.32±2.6% on the Duke-DME. This analysis highlights the importance of a tailored loss weighting scheme that ensures both high-fidelity OCT image synthesis and precise anatomical localization, as confirmed by the PostMask Dice score.

## *D. Cross-Task Analysis: Relationship Between the Translation and Prediction Tasks*

As observed in Tables II and III, SCOD-GAN demonstrates SOTA performance across both tasks, highlighting its superior neural architecture design with acceptable inference time. The model delivers high-quality translations that preserve and/or transform essential clinical features and patterns (e.g., image texture, retinal inner structures, and central thickness), directly contributing to improved classification performance. Additionally, SCOD-GAN showcases remarkable robustness to domain shifts, excelling across different datasets, populations, and conditions. This capability explains its superior prediction scores, even under challenging conditions such as the 10% threshold and external validation. In contrast, models with weaker translation quality struggle in classification due to noise or inconsistencies in preserving and/or translating structural details for post-treatment OCT images. The reduced quality of translated images and unrealistic post-treatment image feature generation negatively impact OCT data image representation, resulting in lower treatment response prediction scores. Furthermore, metrics from Table II—particularly SSIM, FID, and KID—serve as strong predictors of the anti-VEGF treatment response prediction performance.

### *E. Failure Case Discussion*

To further explore the relationship between PostOCT image generation and treatment response prediction tasks, an in-depth analysis was conducted, focusing on the model's failure modes. This analysis reveals a critical interplay between the correctness of synthesis and classification accuracy, particularly in clinically ambiguous scenarios. As illustrated in Fig. 6, we identified a pronounced failure rate for borderline cases, characterized by a marginal post-treatment change in CST within an approximate ±5% range. For this specific patient cohort, prediction errors systematically correspond with anatomical inaccuracies in the synthesized OCT images. We hypothesize that this localized performance drop is a direct consequence of data sparsity. These borderline cases are significantly underrepresented in the training distribution, leading to an inadequately learned decision boundary for subtle pathological changes. This finding underscores the model's sensitivity to class imbalance and highlights a crucial challenge for clinical translation: achieving robust and reliable performance across the full spectrum of disease presentation, especially for cases that lack a definitive therapeutic response.

### *F. Benchmark Against Direct Classification*

To assess the potential performance trade-offs associated with the generative formulation relative to direct response prediction, we compared SCOD-GAN with CADNet, the leading direct classification model reported by Rasti et al. [4], using the identical 5-fold CV protocol and a 10% CST-reduction threshold on the Duke-DME dataset. Crucially, the CADNet framework is an established, highly competitive benchmark previously shown to outperform a comprehensive spectrum of standard ML and DL baselines, including feature-based Sparse Logistic Regression and Extra Trees, as well as prominent CNNs such as VGG16, GoogLeNet, InceptionV3, ResNet50, and Xception. CADNet achieved an Acc of 80.31±5.6%, Pr of 83.96±5.1%, Se of 76.43±7.4%, and F1-score of 80.02±7.0%, whereas SCOD-GAN attained 85.69±4.9%, 89.76±4.6%, 75.97±6.1%, and 82.29±4.7%, respectively. Although CADNet yielded a slightly higher Se, SCOD-GAN demonstrated superior Acc, Pr, and F1-score, excelling in three of the four evaluated classification metrics. SCOD-GAN also achieved a higher case-level AUC than CADNet (0.879±0.06 vs. 0.851±0.08). Beyond binary response classification, SCOD-GAN also generates a structural representation of the predicted post-treatment retina (PostMask) and synthetic post-treatment OCT B-scans (PostOCT), enabling anatomical inspection of the predicted treatment-related changes.

### *G. Summary of Methodological Novelty and Contributions*

From a technical standpoint, the core contribution of our work lies in the sophisticated architectural and methodological integration embodied in SCOD-GAN. The framework is specifically designed to overcome the inherent trade-offs between diffusion models and GANs: by embedding a diffusion process within a GAN architecture, we combine the training stability and sample diversity of diffusion models with the sampling efficiency of adversarial learning, enabling generation in as few as 4 steps. This accelerated hybrid generator is further embedded within a novel, end-to-end optimized cascaded pipeline tailored for clinical prediction. This design also distinguishes SCOD-GAN from SynDiff [28], the seminal adversarial diffusion model for unpaired cross-modality translation. Whereas SynDiff conditions its diffusive modules on a single source-domain image within an unsupervised cycle-consistent framework, SCOD-GAN is trained end-to-end on paired longitudinal OCTs to model treatment-induced anatomical change, routes the translation through an intermediate thickness-mask bottleneck with BCE-Dice supervision, and dual-conditions its second stage on both the baseline OCT and the synthesized mask to enable downstream response prediction. Rather than performing direct and unconstrained translation, SCOD-GAN first synthesizes an intermediate anatomical mask—a decomposition strategy that simplifies the task and enables targeted supervision. This mask is optimized via a hybrid loss function to ensure anatomical fidelity and is then used as a clinically meaningful prior to guide the generation of high-fidelity post-treatment OCT scans.

In addition, SCOD-GAN introduces retinal thickness-aware conditioning, allowing the model to better capture inter-patient anatomical variability and to more precisely model treatment-induced changes in retinal structure. This two-stage generation process—driven by anatomical insight and optimized through compound loss functions—enhances both the accuracy and clinical interpretability of the predicted outcomes. Overall, the holistic design of SCOD-GAN, which integrates rapid hybrid generation, anatomical supervision, and retinal-thickness awareness, represents a significant advancement in OCT-based modeling of therapeutic response.

### *H. Limitations and Future Directions*

Despite promising results, this study has certain constraints. Enhancing image resolution, cross-domain generalization, and robustness against borderline cases requires larger, balanced multi-center longitudinal cohorts. Future pre-processing pipelines could also benefit from full retinal flattening to standardize input distributions. Clinically, because incomplete visual acuity records in the Duke-DME dataset restricted our endpoint to CST reduction, subsequent studies should pair functional vision metrics and patient metadata with multi-layer retinal segmentation for a finer anatomical and functional evaluation. Methodologically, with the availability of larger multi-center longitudinal OCT datasets, scaling the framework to 3D diffusion-GANs and investigating Transformer-based image translation backbones present interesting perspectives for future research. Additionally, integrating explicit treatment-response priors and classification-driven loss terms into the optimization objective can be explored to better guide the synthesis of responder- and non-responder-specific OCT scans.

## VI. Conclusion

In this study, we presented the first thickness-aware source-conditional adversarial generative technique for OCT image translation that efficiently predicts therapeutic response in DME patients treated with monthly anti-VEGF injections. The SCOD-GAN framework leverages a denoising diffusion GAN approach, with which the reverse process of the diffusion model is parametrized by conditional GANs. This model holds promise for optimizing treatment selection for DME patients, potentially reducing unnecessary procedures and improving long-term visual outcomes. Our research aligns with the growing interest in utilizing deep learning methods for personalized medicine and optimizing treatment strategies.

# Supplementary Material

## A Novel Conditional Adversarial Diffusion Model for High-Fidelity OCT-to-OCT Image Translation and Precise Anti-VEGF Therapy Response Prediction in Diabetic Macular Edema

Reza Rasti, Member, IEEE, Armin Biglari, Sobhan Mohammadi Hafshejani, Alireza Dehghani, Michael J. Allingham, and Sina Farsiu, Fellow, IEEE

## Section S1: Quantitative Analysis of Inter-Cohort Distributional Shifts

### A. Multi-Center Cohort Profiles and Acquisition Specifications

To quantitatively contextualize the domain shift between the development cohort and the independent external validation cohort, detailed demographic, clinical, pharmacological, and scanner acquisition characteristics are summarized in Table S1.

Table S1. Comparative Demographic, Clinical, and Acquisition Characteristics of Duke-DME and Isfahan-DME Cohorts

| Characteristic / Parameter | Duke-DME Cohort (Development & 5-Fold CV) | Isfahan-DME Cohort (External Validation) | Comparison / Domain Difference |
|---|---|---|---|
| **Total Patient Cases (n)** | 127 | 44 | Independent external cohort |
| **Geographic Origin / Center** | Duke Eye Center, Durham, NC, USA | Feiz Hospital, Isfahan, Iran | Geographic and clinical domain shift |
| **Population background** | Multi-ethnic U.S. population (White, Black/African American, Asian, Native Hawaiian/Other Pacific Islander, and Unknown) | Iranian population | Population-background difference |
| **Age (years, Mean ± SD)** | N/A | 55–79 (66.2 ± 6.1) | Direct comparison not possible |
| **Sex (Female / Male, n [%])** | N/A | 24 / 20 (54.5% / 45.5%) | Direct comparison not possible |
| **Eye (OD / OS, n[%])** | 70 / 57 (55.1% / 44.9%) | 28 / 16 (63.6% / 36.4%) | Comparable eye laterality distributions |
| **Baseline CST (range (µm), Mean ± SD)** | 172–789 (443 ± 135) | 247–780 (407 ± 138) | Different baseline thickness distributions |
| **Pharmacological Agent (n [%])** | | | |
| **— Bevacizumab (Avastin)** | 69 (54.3%) | 20 (45.5%) | Drug heterogeneity |
| **— Aflibercept (Eylea)** | 44 (34.6%) | 24 (54.5%) | Different agent proportions |
| **— Ranibizumab (Lucentis)** | 14 (11.1%) | 0 (0%) | Agent absent from Isfahan cohort |
| **Treatment Regimen (Injections, n [%])** | | | |
| **— 2 Injections** | 0 (0%) | 12 (27.3%) | Protocol variation |
| **— 3 Injections** | 127 (100%) | 27 (61.4%) | Protocol variation |
| **— 4 Injections** | 0 (0%) | 5 (11.4%) | Protocol variation |
| **Total number of injections, mean ± SD** | 3.00 ± 0.00 | 2.84 ± 0.61 | Different treatment exposure |
| **Injection Interval** | 4 to 6 weeks | 4 to 6 weeks | Standard clinical care |
| **OCT Acquisition Characteristics** | | | |
| **OCT device manufacturer** | Heidelberg Engineering | Heidelberg Engineering | Same manufacturer |
| **OCT system** | Spectralis OCT | Spectralis OCT | Same system family |
| **Acquisition mode** | Spectral-domain OCT (SD-OCT) | Spectral-domain OCT (SD-OCT), High Speed (HS) mode | Acquisition-mode variation |
| **Number of B-scans per volume** | 61 B-scans | 19, 25, or 31 B-scans; (for 9, 23, and 12 cases, respectively) | Different scan density |
| **B-scan resolution** | 768 × 496 pixels | 768 × 496 pixels (most sessions); 512 × 496 pixels (a minority) | Resolution variation in a minority of cases |
| **Axial resolution** | 3.8 µm | 3.8 µm | Same nominal axial resolution |
| **Scan area** | ≈7.8–9.9 × 6.6–8.4 mm (30° field) | ≈8.3–9.3 × 4.6–7.4 mm (30° field); ≈5.7–5.9 mm width (20° field) | Scan geometry variation |

Differences in patient profiles, treatment regimens, scan density, and image dimensions may contribute to cross-center distributional shifts and provide context for interpreting the external validation performance gap.

**B. Feature-Space Latent Distribution Analysis (t-SNE)**

To characterize the feature-space distributional differences between the primary development cohort (Duke-DME) and the external validation cohort (Isfahan-DME), pre-treatment OCT B-scans were encoded using a shared, frozen ImageNet-pretrained ResNet-50, yielding a 2,048-dimensional feature vector per image. To account for differences in scan density between the datasets (61 B-scans per volume in Duke-DME and 19, 25, or 31 in Isfahan-DME), five B-scans per patient were selected at matched relative positions within each volume (0.3–0.7), corresponding approximately to the central 2–3 mm of the macular scan in both datasets. Forty-four patients were sampled from each cohort. Both datasets underwent identical preprocessing: the central 512 columns were retained, and the images were resized to 256 × 256 pixels. The extracted features were reduced to 50 dimensions using principal component analysis (PCA) and subsequently projected into two dimensions using t-distributed stochastic neighbor embedding (t-SNE; perplexity = 30, fixed random seed). The resulting feature distributions are shown in Fig. S1. As a control, Duke-DME was divided into two disjoint, randomly selected patient subsets, which were processed using the same feature extraction, preprocessing, and dimensionality-reduction pipeline.

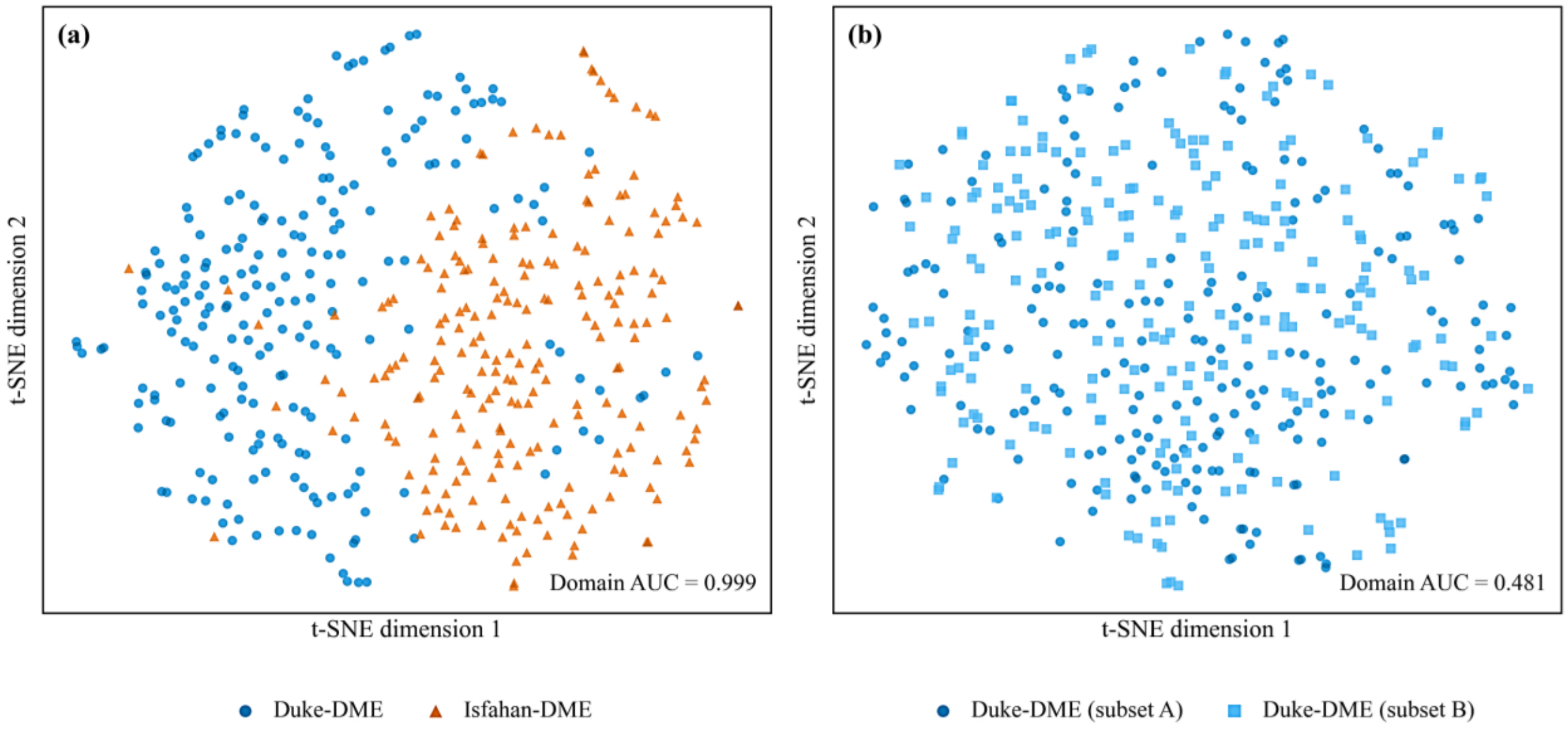

Figure S1. t-SNE feature-space projection of Duke-DME (development) and Isfahan-DME (external validation) representations. The distinct clustering illustrates domain shift, empirically justifying the zero-shot performance drop on the external cohort.

In Fig. S1(a), Duke-DME and Isfahan-DME feature representations occupy largely separate manifolds, with only a narrow central interface where representations overlap. In the negative control (Fig. S1(b)), disjoint Duke-DME subsets are thoroughly intermixed, confirming that the cross-cohort separation is not an artifact of patient sampling, B-scan slice location, or the t-SNE projection itself. Because t-SNE embeddings are qualitative and sensitive to local density, we substantiated this discrepancy quantitatively. A patient-wise domain classifier separated the cohorts with an AUC of 0.999 ± 0.001 (versus 0.48 ± 0.07 in the control); the Fréchet feature distance was 11.7 (versus 4.0 in the control); and only 14% of a scan's ten nearest neighbors belonged to the opposite cohort, compared to 42% under random splitting (50% representing complete mixing). This structural separation remained robust across image smoothing, per-image intensity standardization, and variable padding retention. Furthermore, Isfahan-DME B-scans exhibited an approximately 2.6-fold higher background speckle noise level (median standard deviation of 21.5 vs. 8.2 gray levels). Collectively, these findings demonstrate a pronounced cross-center distribution shift; differences in acquisition mode, sampling density, noise characteristics, and cohort demographics may contribute to this shift, highlighting the domain gap that any translation or predictive model must bridge.

## Section S2: Residual Error Maps for Post-Treatment OCT Synthesis

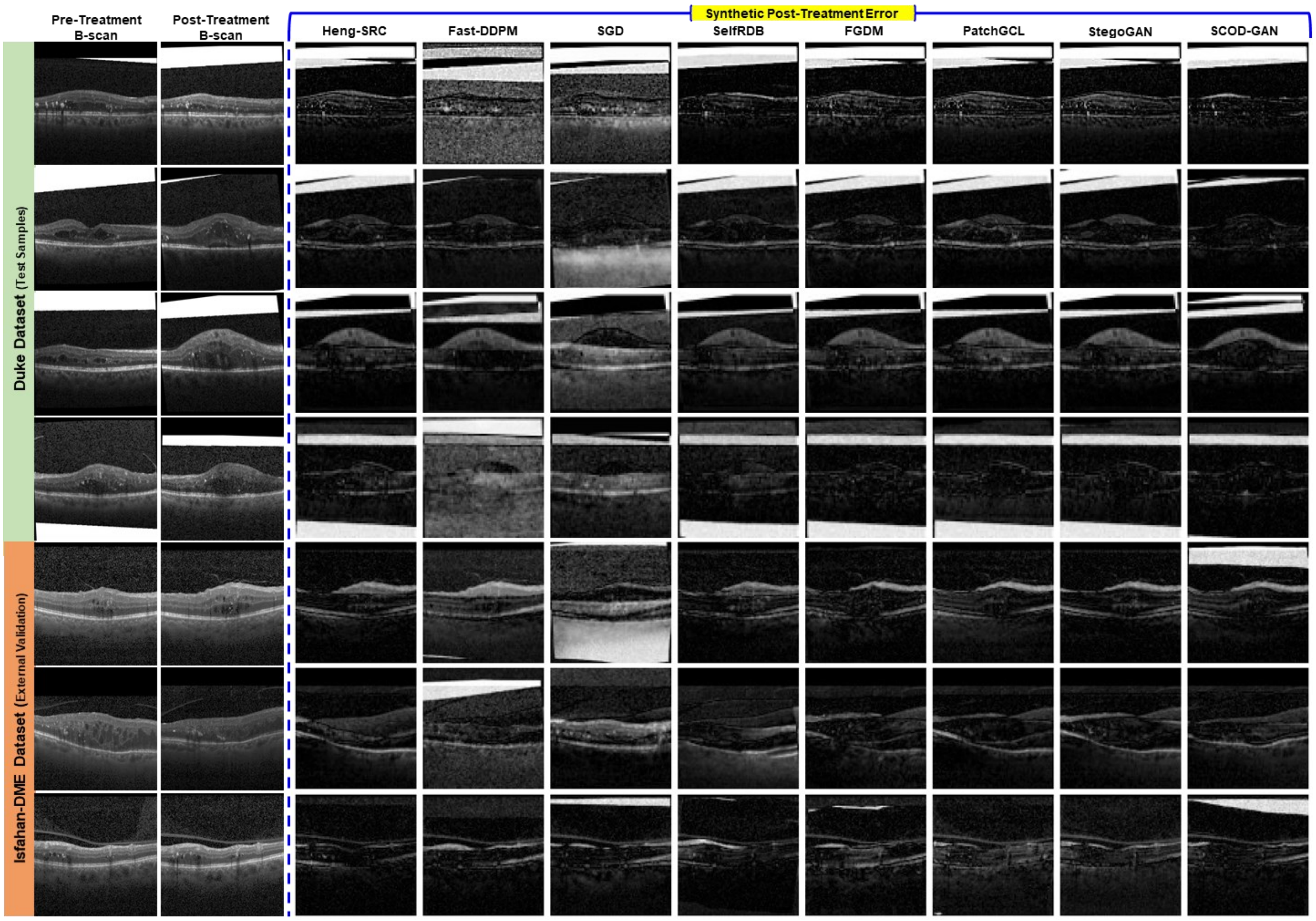

Fig. S2. Pixel-wise absolute error maps for post-treatment OCT B-scan generation by SCOD-GAN and competing methods. The maps show the differences between generated and ground-truth post-treatment OCT scans across retinal morphology and pathological features, including fluid regions, retinal layer curvature, and hyperreflective foci; SCOD-GAN shows comparatively lower errors in these structurally and clinically relevant regions.